\documentclass[iop]{emulateapj}

\usepackage{natbib}
\usepackage{apjfonts}
\usepackage[colorlinks=true,citecolor=blue]{hyperref}
\usepackage{graphicx}
\usepackage[usenames,dvipsnames]{color}

\newcommand{\fuvr}{FUV$-r$\,\,}

\newcommand{\halpha}{H$\alpha$\,\,}
\newcommand{\hst}{\emph{HST}\ }
\newcommand{\esf}{ESF-ETGs}
\newcommand{\sr}{SR2010}

\shorttitle{EXTENDED STAR FORMATION IN GREEN VALLEY GALAXIES}
\shortauthors{FANG ET AL.}

\begin{document}

\title{The Slow Death (or Rebirth?) of Extended Star Formation in $\lowercase{z}\sim0.1$ Green Valley Early-Type Galaxies}

\author{Jerome J.~Fang\altaffilmark{1}, S.~M.~Faber\altaffilmark{1}, Samir Salim\altaffilmark{2}, Genevieve J.~Graves\altaffilmark{3}, and R.~Michael Rich\altaffilmark{4}}

\altaffiltext{1}{UCO/Lick Observatory, Department of Astronomy and Astrophysics, University of California, Santa Cruz, CA 95064, USA; {jjfang@ucolick.org}}
\altaffiltext{2}{Department of Astronomy, Indiana University, Bloomington, IN 47404, USA}
\altaffiltext{3}{Miller Fellow, Department of Astronomy, University of California, Berkeley, CA 94720, USA}
\altaffiltext{4}{Department of Physics and Astronomy, University of California, Los Angeles, CA 90095, USA}

\submitted{ApJ Accepted, 2012 October 17}

\begin{abstract}

UV observations in the local universe have uncovered a population of early-type galaxies with UV flux consistent with low-level recent or ongoing star formation.  Understanding the origin of such star formation remains an open issue. We present resolved UV-optical photometry of a sample of 19 SDSS early-type galaxies at $z\sim0.1$ drawn from the sample originally selected by \citet{salim10} to lie in the bluer part of the green valley in the UV-optical color-magnitude diagram as measured by GALEX. Utilizing high-resolution \hst far-UV imaging provides unique insight into the distribution of UV light in these galaxies, which we call ``extended star-forming early-type galaxies'' (\esf) because of extended UV emission that is indicative of recent star formation. The UV-optical color profiles of all \esf\ show red centers and blue outer parts. Their outer colors require the existence of a significant underlying population of older stars in the UV-bright regions. Analysis of stacked SDSS spectra reveals weak LINER-like emission in their centers. Using a cross-matched SDSS DR7/GALEX GR6 catalog, we search for other green valley galaxies with similar properties to these \esf\ and estimate that $\approx13\%$ of dust-corrected green valley galaxies of similar stellar mass and UV-optical color are likely ESF-candidates, i.e., \esf\ are not rare. Our results are consistent with star formation that is gradually declining in existing disks, i.e., the \esf\ are evolving onto the red sequence for the first time, or with rejuvenated star formation due to accreted gas in older disks provided that the gas does not disrupt the structure of the galaxy and the resulting star formation is not too recent and bursty. \esf\ may typify an important subpopulation of galaxies that can linger in the green valley for up to several Gyrs, based on their resemblance to nearby gas-rich green valley galaxies with low-level ongoing star formation. 

\end{abstract}


\section{INTRODUCTION} \label{introduction}

Early-type galaxies (ETGs) are no longer viewed as a homogeneous population of ``red and dead'' systems that have ceased forming stars at late times. A mounting wealth of evidence indicates that some ETGs have experienced recent or ongoing star formation (SF). In the nearby universe, the SAURON survey \citep{dezeeuw02} has uncovered recent SF in a subset of the ETGs in its sample. (Note that the SAURON galaxies are only a representative sample of local ETGs rather than a complete, volume-limited sample.) \citet{shapiro10} found that the $\sim30\%$ of SAURON galaxies with infrared signatures of SF (i.e., PAH emission) are preferentially classified as ``fast rotators'' (galaxies harboring an embedded disc component), while the non-star forming galaxies are ``slow rotators'' (spheroidal, kinematically hot systems). This result was corroborated by a stellar population analysis of absorption line maps of the SAURON galaxies: a subset of the fast rotators have SSP-equivalent ages consistent with recent SF, while slow rotators have uniformly old ages \citep{kuntschner10}. Intriguingly, the fast rotators tend to be morphologically classified as S0 (lenticular) galaxies, while the slow rotators are almost exclusively E (elliptical) galaxies \citep{emsellem07}. This suggests a possible morphological dependence on the observed SF in ETGs, with recent SF preferentially occurring in S0 galaxies.

Evidence for recent or ongoing SF in ETGs is also found in much larger samples available from all-sky surveys, including the Sloan Digital Sky Survey \citep[SDSS;][]{stoughton02}, and the Galaxy Evolution Explorer \citep[GALEX;][]{martin05}. The latter provides UV measurements that are crucial in constraining recent SF \citep{kaviraj09}. Utilizing both SDSS and GALEX photometry, \citet{yi05} showed that at least $\sim15\%$ of their sample of 160 nearby ($z\la0.1$) ETGs have blue UV-optical colors that they argue can only be attributed to low-level recent SF. With a larger sample of $\sim2100$ morphologically selected SDSS ETGs, \citet{kaviraj07} concluded that $\sim30\%$ of their sample have UV-optical colors consistent with SF occurring within the past 1 Gyr, forming $\sim1\%$--$3\%$ of the total stellar mass.

Given these results, it is clear that recent SF in ETGs is more prevalent than once believed. What remains uncertain is how to interpret such SF in the context of galaxy evolution. As is well known, galaxies exhibit a bimodal color distribution, with star-forming galaxies located in the ``blue cloud'' and passively evolving objects located in the ``red sequence'' \citep[e.g.,][]{strateva01}. In between the blue cloud and red sequence lies the ``green valley'' (GV), which is most apparent when studying UV-optical colors. The movement of galaxies through the GV is presumed to be a net flow from blue to red as a result of the quenching of SF in blue cloud galaxies and the resultant reddening of their colors \citep{bell04,faber07,martin07}. Within this framework, the ETGs with recent SF are a mystery: is the observed SF the most recent phase of the original (fading?) star formation, or is it a subsequent episode that has been triggered in a hitherto quiescent galaxy? 

Much attention has been focused on the latter explanation of the recent SF in ETGs, which has been termed ``rejuvenation'' in the literature \citep[e.g.,][]{rampazzo07,thomas10,thilker10,marino11}. The newly triggered SF could drive a galaxy \emph{back into} the GV from the (UV-optical) red sequence. In this paper, ``rejuvenation'' is used to describe SF that causes a quiescent red sequence galaxy to move back into the GV. 

One possible cause of rejuvenated SF is gas accretion during gas-rich minor mergers \citep[e.g.,][]{kaviraj09}. Recently, \citet{thilker10} showed that the SF observed by GALEX in the outer regions of the nearby S0 galaxy NGC 404 is most likely fueled by an external reservoir of \ion{H}{1} gas formed during a recent merger with a gas-rich dwarf galaxy \citep{delrio04}. Gas-rich major mergers can also trigger nuclear SF; however, the infalling gas is also predicted to trigger strong AGN feedback that expels or heats up the remaining cold gas, effectively quenching SF \citep[e.g.,][]{dimatteo05,hopkins06}.  

Another possible trigger for rejuvenated SF is smooth accretion from the intergalactic medium (IGM). Direct evidence for such accretion has been difficult to observe. However, the presence of extended reservoirs of neutral gas around ETGs has been shown to be a fairly common phenomenon \citep[e.g.,][]{morganti06,catinella10}. The formation of some of these \ion{H}{1} disks has been attributed to accretion from the IGM \citep[e.g.,][]{finkelman11,wang11}. \citet{cortese09} found that the ``\ion{H}{1}-normal'' galaxies in their sample of GV galaxies are experiencing recent SF that in a few cases may be rejuvenated, owing to evidence for recently accreted \ion{H}{1} disks (i.e., the \ion{H}{1} disks show disturbed or ring-like morphologies indicative of external accretion).

It is important to note that the UV flux seen in ETGs is not necessarily due to recent SF. An excess of UV flux seen in elliptical galaxies (the ``UV upturn'') is believed to be caused by hot, evolved stars \citep[e.g., horizontal branch stars;][]{code69,oconnell99}. In this case, the UV emission is expected to be smoothly distributed and follow the optical light profile of the galaxy. Note, however, that UV upturn ETGs are found in the UV-optical red sequence and not in the GV \citep{yi05}. AGN activity can also contribute to the UV light seen in ETGs \citep[e.g.,][]{agueros05}.

Our understanding of recent SF and rejuvenation in ETGs has been drastically improved thanks to the large samples provided by SDSS and GALEX. However, due to the limited spatial resolution of GALEX and the typical redshift of SDSS galaxies ($z\sim0.1$), it has not been possible to resolve the actual physical location of the SF that is inferred based on integrated UV-optical colors. Combining high-resolution UV imaging with the large sample sizes of all-sky surveys could provide important information about the nature and origin of recent SF in ETGs.

As a follow-up to the SDSS/GALEX studies mentioned above, \citet[][hereafter \sr]{salim10} reported first results from high-resolution \hst UV imaging of a sample of 29 SDSS ETGs detected by GALEX. These galaxies were chosen to have high SDSS concentration and blue UV-optical colors, yet no detectable emission lines in the 3\arcsec\ SDSS fiber spectra. This unique combination was intended to select centrally quiescent ETGs yet with high UV excess. The emission line cut also had the effect of discarding ETGs with emission lines due to AGN activity, including most LINERs.

\sr\ found that $\sim2/3$ of their sample show \emph{extended} UV emission in rings, clumps, and/or spiral arms due to ongoing or recent SF. In \citet[][hereafter Paper I]{salim12}, these galaxies were named ``extended star-forming early-type galaxies'' (\esf), and we do the same in this work. The remaining $\sim1/3$ of their sample had only small-scale SF or centrally compact\footnote{These compact-UV objects are referred to as ``unresolved'' UV objects in Paper I.} UV emission. According to \sr, the \esf\ on average have fairly low far-UV (FUV) dust attenuation and large UV sizes with respect to other galaxies with similar specific star formation rates. The large UV sizes and low FUV dust attenuation were consistent, they argued, with rejuvenated SF taking place in large outer reservoirs of gas acquired in mergers with gas-rich dwarfs or directly from the IGM. However, fading of the original SF could not be ruled out in at least part of the ESF-ETG sample.

A more comprehensive exploration of the same sample is presented in Paper I, supplemented with deep, good-seeing $R$-band images taken with the WIYN telescope. The discussion there is focused on the qualitative morphologies of the \esf\ as determined from optical and UV images. Galaxies were classified according to their optical Hubble types as well as by their UV morphologies. Some key conclusions from Paper I are (1) the \esf\ have smooth optical morphologies without signatures of recent merger activity, (2) the origin of the gas is either internal or smoothly accreted from the IGM, and (3) the extended SF phenomenon is preferentially found in S0s and not ellipticals.

The intriguing results discussed in \sr\ and Paper I call out for additional data and analysis to provide further insight into the nature of \esf\ in the GV. Quantitative photometry from the optical and UV images can provide additional information about the stellar populations in the \esf. In addition, a more careful comparison between the \esf\ and the SDSS/GALEX parent population from which they were selected can help elucidate whether these objects are rare or common. To tackle these issues, we take advantage of the available UV and optical imaging to present a quantitative analysis of the star formation \emph{within} each galaxy via aperture photometry and stellar population modeling. We show that the \esf\ have characteristically red centers containing old stars and blue outer disks due to recent SF, which is expected given how they were selected. The outer disks have a significant population of old stars as well, and we show that the colors are consistent with either the gradual decline of SF or rejuvenation occurring on extended timescales. In addition, we show that a sizable fraction of massive GV galaxies are potentially \esf.

The paper is structured as follows. We describe the \sr\ sample of 29 ETGs, sources of data, and data reduction in Section \ref{data}. Section \ref{images} presents optical and UV photometry of the sample, revealing similarities among the \esf. An analysis of the stellar populations in the outer disks of the \esf\ is presented in Section \ref{sps_ext}. Section \ref{sect_spec} presents stacked SDSS spectra of the bulges of the \esf, revealing old central ages and hints of weak emission activity with LINER-like ratios. We search for additional candidate \esf\ from a large sample of SDSS galaxies and estimate the ESF-ETG contribution to the GV in Section \ref{sect_gv}. An examination of various possible scenarios to explain the SF in the \esf\ is given in Section \ref{discussion}. Our main conclusions are summarized in Section \ref{conclusions}. 

All magnitudes in this paper are on the AB system \citep{oke74}. A concordance $\Lambda$CDM cosmology with $\Omega_m=0.3,\Omega_\Lambda=0.7,$ and $H_0=70\mathrm{\,km\,s^{-1}\,Mpc^{-1}}$ is assumed.

\section{DATA AND REDUCTION} \label{data}

Descriptions of the sample selection criteria used in \sr\ as well as specifics about the UV and optical imaging are discussed in detail in Paper I. Briefly, a sample of 30 quiescent ETGs with strong UV excess (\fuvr$<5.3$) were selected from a cross-matched SDSS DR4/GALEX IR1.1 catalog \citep{seibert05}. High-resolution \hst Solar-Blind Channel FUV images were obtained for 29 objects. The numbering scheme used in Paper I to identify these objects is retained in this paper, e.g., SR01 is the first object in the sample. Objects are ordered by increasing \fuvr color.

As discussed in Paper I, two objects (SR22 and SR26) were removed from the analysis because of incorrect GALEX IR1.1 FUV magnitudes that resulted in erroneously blue \fuvr colors. It was shown that the remaining galaxies divide nicely into three categories based on their UV morphology: extended SF (UV rings and/or arms, 19 objects), small-scale SF (6 objects), and compact (central UV source only, 2 objects). Because the \esf\ represent the majority of the sample and may furthermore represent the behavior of many massive GV galaxies, we focus exclusively on them in the rest of this paper.

\subsection{SDSS and GALEX Data} \label{sect_ancillary}

Placing the \esf\ into the broader context of evolution through the GV requires a broad suite of ancillary data. In particular, we take advantage of additional structural and spectral data available from the SDSS and GALEX databases. To make use of updated data since \sr, we used SDSS DR7 \citep{abazajian09} and GALEX GR6 for both the \esf\ and a larger SDSS sample. This larger sample was constructed using the SDSS DR7/GALEX GR6 cross-matched catalog available through the GALEX CASJobs interface\footnote{\texttt{http://galex.stsci.edu/casjobs/}}. SDSS galaxies with redshifts $0.005<z<0.12$ having a GALEX NUV and/or FUV detection in the Medium Imaging Survey (MIS) within a 5\arcsec\,radius were selected. Objects were excluded if they were more than 0\fdg55 from the center of the GALEX detector to ensure reliable photometry. The resulting $\sim57,000$ galaxies are referred to in this paper as the ``SDSS master sample.'' In Section \ref{sect_gv}, we select galaxies from this sample and compare their properties with the 19 \emph{HST} \esf.

For both the \esf\ and the SDSS master sample, integrated magnitudes in GALEX FUV, NUV, and SDSS $u,g,r,i,\mathrm{and}\ z$ were retrieved. Spectroscopic redshifts, Galactic extinction, optical Petrosian radii (enclosing 50\% of the Petrosian flux), isophotal axis ratios (\texttt{isob\_r/isoa\_r}), and GALEX FWHM measurements (a measure of UV diameter) were also collected; the GALEX resolution is $\approx5\arcsec$ FWHM. All integrated magnitudes were corrected for Galactic extinction and $k$-corrected to $z=0$ using the \texttt{kcorrect} code, version 4.2 \citep{blanton07}. For reference, the median $k$-correction for blue cloud (red sequence) galaxies is $\Delta($FUV$-r)=0.08\ (0.17)$ and $\Delta(g-r)=0.06\ (0.17)$, where $\Delta(\mathrm{color})$ is the difference between observed and rest-frame color. Stellar mass estimates and emission line measurements were obtained from the MPA/JHU value-added catalogs for SDSS DR7\footnote{A Chabrier IMF was assumed in calculating stellar masses. Catalogs are available at \texttt{http://www.mpa-garching.mpg.de/SDSS/DR7}}. Table \ref{table_prop} lists the various SDSS and GALEX quantities described above for the \esf. We also obtained \emph{ugriz} aperture photometry for the \esf\ from the SDSS DR7 database's {\tt frames} pipeline. These data consist of radially averaged surface brightness profiles measured in a series of circular annuli centered on each object. Aperture colors are corrected for Galactic extinction only, i.e., they have not been $k$-corrected. The optical aperture photometry is used in conjunction with the \emph{HST} UV photometry to measure color profiles (Section \ref{images}). Flux-calibrated SDSS spectra for the \esf\ were also obtained for use in characterizing their central bulges (Section \ref{sect_spec}).

\begin{deluxetable*}{cccccccccc}
\tabletypesize{\footnotesize}
\tablewidth{0in}
\tablecaption{Properties of the \esf \label{table_prop}}

\tablehead{\colhead{Object} & \colhead{Redshift} & \colhead{$\log M_*$} & \colhead{$\ \ \ M_r$} & \colhead{$r$} &
	   \colhead{\fuvr} & \colhead{$E(B-V)$} &  \colhead{FUV FWHM}  & \colhead{\halpha EW} & \colhead{$b/a$}\\	   
           \colhead{} & &\colhead{$[M_\odot]$} & & & & &\colhead{(kpc)}   &  \colhead{(\AA)} & }

\startdata           
SR01 & 0.119 & 10.94& $-21.99$ & 16.72 & 2.98 &   0.059 & 91.87  &    
1.251 & 0.86 \\ 
SR02 & 0.081 & 10.15& $-20.11$ & 17.72 & 3.09 &   0.044 & 57.60  &       0.603 & 0.92 \\ 
SR03 & 0.057 & 10.58& $-20.69$ & 16.35 & 4.34 &   0.016 & 43.19  &        0.200 & 0.61 \\ 
SR04 & 0.114 & 10.64& $-21.34$ & 17.28 & 4.20 &   0.028 & 60.38  &      1.352 & 0.68 \\ 
SR05 & 0.101 & 10.56& $-21.10$ & 17.24 & 4.18 &   0.061 & 23.14  &         2.098 & 0.56 \\ 
SR06 & 0.095 & 10.81& $-21.68$ & 16.51 & 4.33 &   0.040 & 44.57  &         0.587 & 0.56 \\ 
SR07 & 0.107 & 10.73& $-21.44$ & 17.04 & 4.69 &   0.036 & 31.42  &        1.406 & 0.75 \\ 
SR08 & 0.111 & 10.75& $-21.46$ & 17.10 & 4.01 &   0.027 & 57.57  &         0.514 & 0.75 \\ 
SR09 & 0.092 & 10.42& $-20.72$ & 17.39 & 4.93 &   0.042 & 25.58  &        0.645 & 0.57 \\ 
SR10 & 0.111 & 11.22& $-22.51$ & 16.04 & 4.92 &   0.026 & 41.76  &         0.714 & 0.65 \\ 
SR11 & 0.077 & 10.68& $-21.48$ & 16.25 & 4.46 &   0.027 & 47.33  &        0.779 & 0.76 \\ 
SR12 & 0.114 & 11.03& $-22.36$ & 16.26 & 4.83 &   0.024 & 54.43  &          0.852 & 0.57 \\ 
SR14 & 0.085 & 11.02& $-22.00$ & 15.93 & 5.27 &   0.088 & 29.48  &          0.736 & 0.42 \\ 
SR17 & 0.118 & 10.96& $-22.01$ & 16.69 & 5.13 &   0.051 & 38.78  &          2.061 & 0.75 \\ 
SR18 & 0.116 & 10.72& $-21.35$ & 17.31 & 4.88 &   0.022 & 34.67  &          2.232 & 0.46 \\ 
SR20 & 0.114 & 10.87& $-21.66$ & 16.97 & 5.13 &   0.025 & 45.98  &          2.160 & 0.59 \\ 
SR23 & 0.114 & 11.13& $-22.30$ & 16.32 & 5.22 &   0.039 & 55.80  &           0.799 & 0.95 \\ 
SR28 & 0.104 & 10.73& $-21.38$ & 17.02 & 5.38 &   0.033 & 33.82  &          2.326 & 0.48 \\ 
SR29 & 0.059 & 10.38& $-20.73$ & 16.37 & 5.36 &   0.014 & 18.10  &         0.889 & 0.70 

\enddata

\tablecomments{Redshifts, stellar masses, and \halpha equivalent widths from the SDSS DR7 MPA/JHU value-added catalog. Optical photometry and axis ratios ($b/a$) from the SDSS DR7 database. FUV magnitudes, FUV FWHM, and color excess from the GALEX GR6 database. All photometry in table $k$-corrected to $z=0$ and corrected for Galactic extinction using the $E(B-V)$ values listed.}

\end{deluxetable*}

\begin{figure}
	\epsscale{1.2}
	\plotone{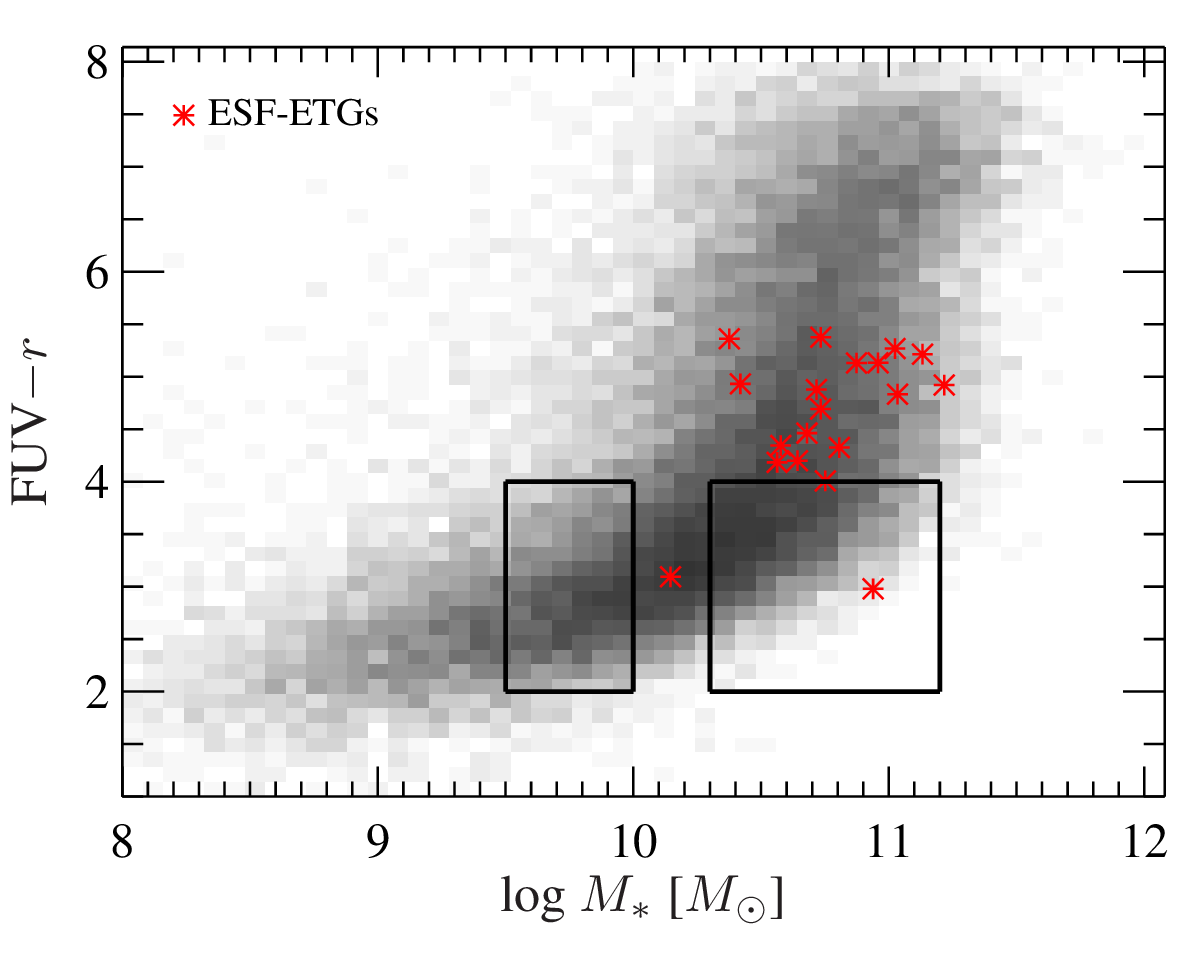}
	
	\caption{\fuvr color-stellar mass diagram, with the SDSS master sample plotted as the gray histogram and the \esf\ plotted as red points. AB magnitudes are used. Stellar masses are from the MPA/JHU SDSS DR7 Value-Added Catalog. The \esf\ are concentrated in the bluer part of the green valley ($4\la\,$\fuvr$\la5.3$) and are relatively massive ($\log M_*/M_\odot\ga10.5$). The rectangles enclose low-mass and high-mass blue galaxies with \emph{integrated} \fuvr colors similar to the \emph{outer} \fuvr colors of the \esf\ $(2<\ $\fuvr$<4)$. Colors of objects within the rectangles are compared to the outer colors of the \esf\ in Figure \ref{models}.}

	\label{fuvr_mass}
\end{figure}

Figure \ref{fuvr_mass} shows the distribution of the \esf\ and the SDSS master sample in \fuvr color vs.~stellar mass. This figure demonstrates how cleanly \fuvr separates the main galaxy population into well-defined blue and red sequences, with a clear GV in between. This is a major advantage of using a UV-optical color; we can easily separate fully quenched objects from those that have only recently quenched. An optical color does not have the dynamic range to do this, which is why the optical red sequence appears more prominent. In other words, the red sequence in Figure \ref{fuvr_mass} appears less well-defined due to the ``resolving power'' of \fuvr color. In addition the figure does not show a volume-limited sample, which reduces the number of low-mass, red objects in the plot. Note that the \esf\ are located among the higher-mass galaxies in the GV (and blue sequence).

\subsection{Aperture Photometry} \label{sect_hstobs}
Aperture photometry of the UV images was measured using Source Extractor, version 2.8.6 \citep{bertin96}. To be consistent with the SDSS optical measurements, UV fluxes were measured using the circular apertures defined by the SDSS \texttt{frames} pipeline. The radii of the apertures used are $R=0\farcs23,0\farcs68,1\farcs03,1\farcs76,3\arcsec,4\farcs63,7\farcs43, \,\mathrm{and}\ 11\farcs42$. While elliptical apertures are better suited for this particular sample, we show below that the typical color profile in the outer parts of the galaxies is fairly flat, and thus our measurements are not that sensitive to the exact shape of the aperture. To check this, we compared the SDSS $r$-band surface brightness profiles to those measured in elliptical apertures from deeper $R$-band images taken with the WIYN telescope (presented in Paper I); no significant difference was found between the shapes of the SDSS and WIYN profiles. The WIYN data are not used here because they exist for only 16 of the 19 \esf\ and because they have not been flux-calibrated. To avoid spurious color gradients, the \hst UV images were first smoothed to match the 1\farcs4 seeing (1$\sigma$) of the SDSS telescope. In a few cases (SR08 and SR17) nearby UV companions located within the outermost aperture were masked out prior to measuring photometry.

The UV sky background in each individual frame was determined by measuring fluxes in a number of $100\times100$-pixel boxes placed in source-free regions of the image. The average value of all the box fluxes was taken to be the sky value and was subtracted from each individual frame prior to co-adding them (Paper I). The sky is not uniform across the image, and this represents an additional source of uncertainty in the photometry. We accounted for this by measuring the rms scatter of the residual sky values in the co-added frames, defined to be the standard deviation of the distribution of box fluxes in source-free regions. The total error in each aperture flux measurement consists of Poisson noise of the source and sky and the rms scatter in the residual sky, all added in quadrature. Note that the SBC detector has no read noise. Flux errors in the optical photometry are small (a few percent at most), and the errors on the colors are dominated by the UV measurements. 

\begin{figure}
	\epsscale{1.2}
	\plotone{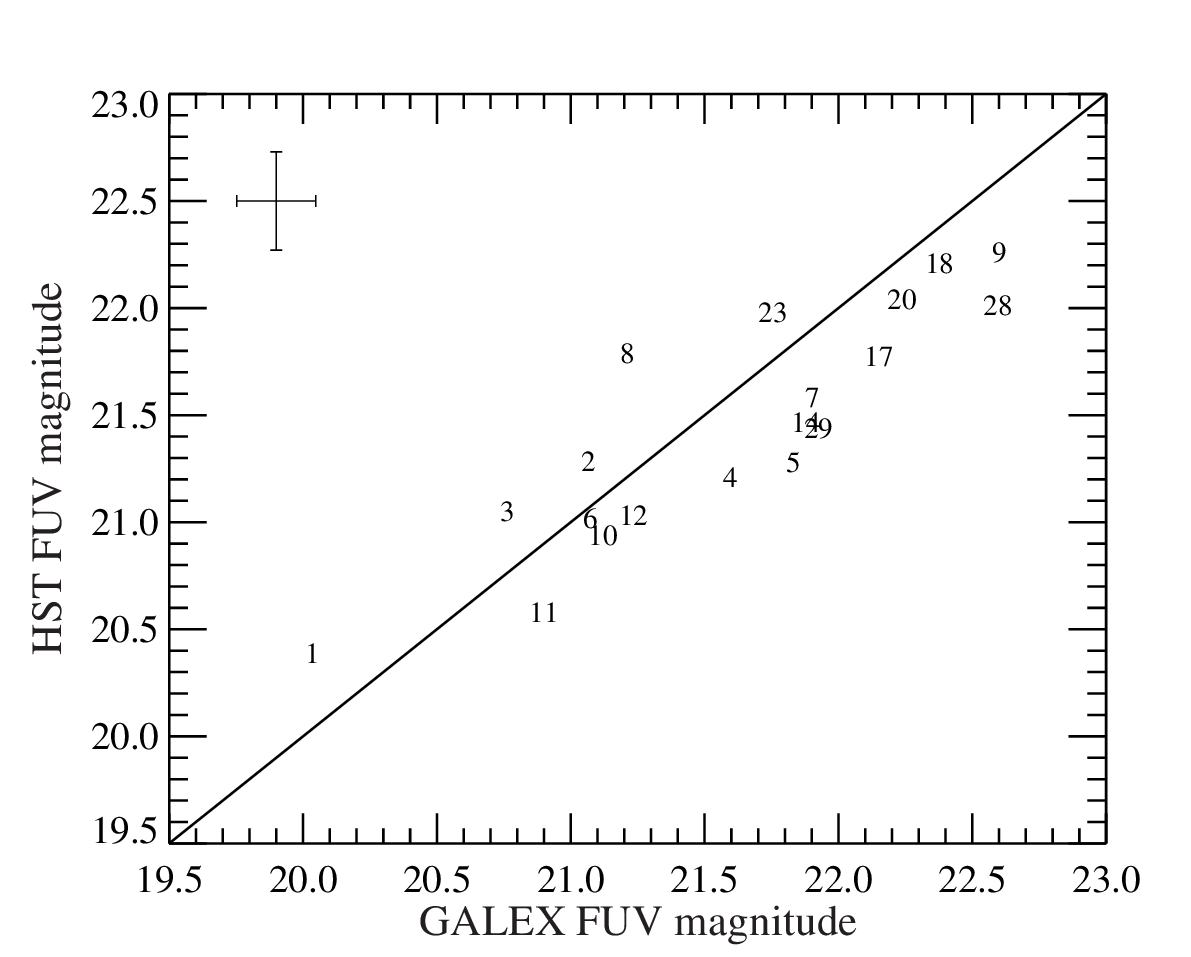}
	
	\caption{Comparison between GALEX FUV ($\lambda_e=1530$\,\AA) and \hst FUV (F125LP filter, $\lambda_e=1430$\,\AA) magnitudes for the \esf. The magnitudes have not been corrected for Galactic extinction. The points are labeled by object number (1 = SR01, etc.). The cross indicates median errors in the magnitudes. The one-to-one line is indicated. The HST magnitudes were measured within a circular aperture of radius $R=11\farcs42$. The agreement between the GALEX and HST magnitudes is generally good, with a median offset of HST$-$GALEX $=-0.19$ mag. Because this is smaller than the \fuvr colors, no zeropoint correction is made to the \hst magnitudes.}

	\label{total_mag}
\end{figure}

To check for any zeropoint offset between the GALEX FUV and \hst FUV filters,\footnote{Our adopted \emph{HST} zeropoint was obtained at \texttt{http://www.stsci.edu/hst/acs/analysis/zeropoints}.} we compared integrated magnitudes measured from the \hst images to the GALEX GR6 ``mag\_auto'' magnitudes. As seen in Figure \ref{total_mag}, the agreement between the two measurements is generally good, with a median offset of HST$-$GALEX $=-0.19$ mag. Since the offset is much smaller than the \fuvr colors, no correction was made to the \hst magnitudes. Note that the magnitudes in Figure \ref{total_mag} have not been corrected for Galactic extinction. 

\section{RESULTS}\label{results}

\subsection{UV-Optical Color Profiles} \label{images}

\begin{figure} 
  \epsscale{1.15}
  \plotone{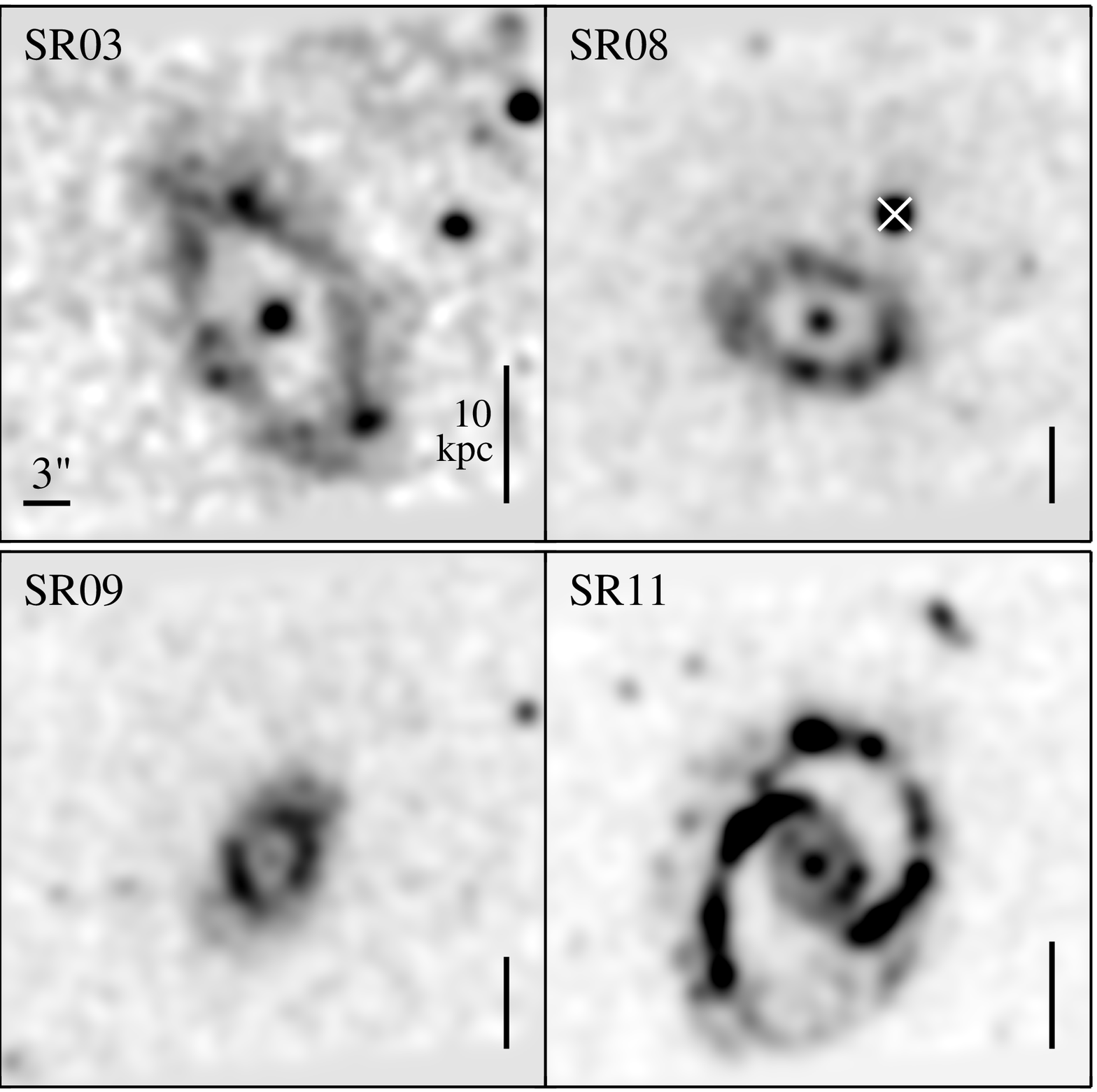}

  \caption{A montage of four \esf\ highlighting the typical UV morphologies observed in these galaxies. The \hst UV images have been smoothed to match the seeing of SDSS and are $\sim30\arcsec$ on a side. For scale, the horizontal bar indicates the 3\arcsec\ diameter of the SDSS spectroscopic fiber, and the vertical bar in each panel equals 10 kpc. The companion marked ``$\times$'' was masked when measuring photometry. Unmarked companions are located outside the maximum aperture used (11\farcs42). Extended UV emission is located \emph{outside} the fiber. The majority of the 19 \esf\ in the sample have similar UV rings. SR11 is one of two objects with pronounced spiral arms. High-resolution images of all 29 \hst UV galaxies are presented in Paper I, along with SDSS optical and WIYN $R$-band images.}
  
  \label{plot_images}
\end{figure}

To give an idea of the UV morphologies seen in the \esf, Figure \ref{plot_images} presents a montage of four \esf\ in the sample. As discussed in Paper I, UV rings are seen in all but one of the \esf, and clear spiral arms are found in two objects. Crucially, the UV emission is \emph{extended} over tens of kpc in all cases. Compact, central UV emission is also seen in nearly all the \esf. The reader is referred to Paper I for a compilation and discussion of the UV and optical images and morphologies of the complete sample.

New in this paper are the SDSS $r$-band and \hst UV surface brightness and color profiles, presented in Figure \ref{plot_prof}. Surface brightness measurements are also listed in Table \ref{table_sb} for reference (negative fluxes are arbitrarily assigned a surface brightness of 35 mag arcsec$^{-2}$). Since the surface brightness values are averages within each annulus, each point has been plotted at the midpoint of each annulus. To facilitate comparison between galaxies, the radial coordinate has been scaled to the Petrosian half-light radius ($R_{50}$) for each galaxy. The photometry quantitatively highlights the disconnect between optical and UV light in tracing young stars and underscores the leverage one gains with UV data when studying recent SF in ETGs. Compared to the $r$-band surface brightness profiles, the UV profiles are much shallower in general, particularly in the outer parts. In principle, one could measure UV surface brightnesses to larger radii to investigate where the UV light falls off. However, larger apertures would often reach beyond the edge of the detector, making accurate sky measurements difficult, especially for the larger galaxies.

\begin{figure*} 
  \epsscale{1.0}
  \plotone{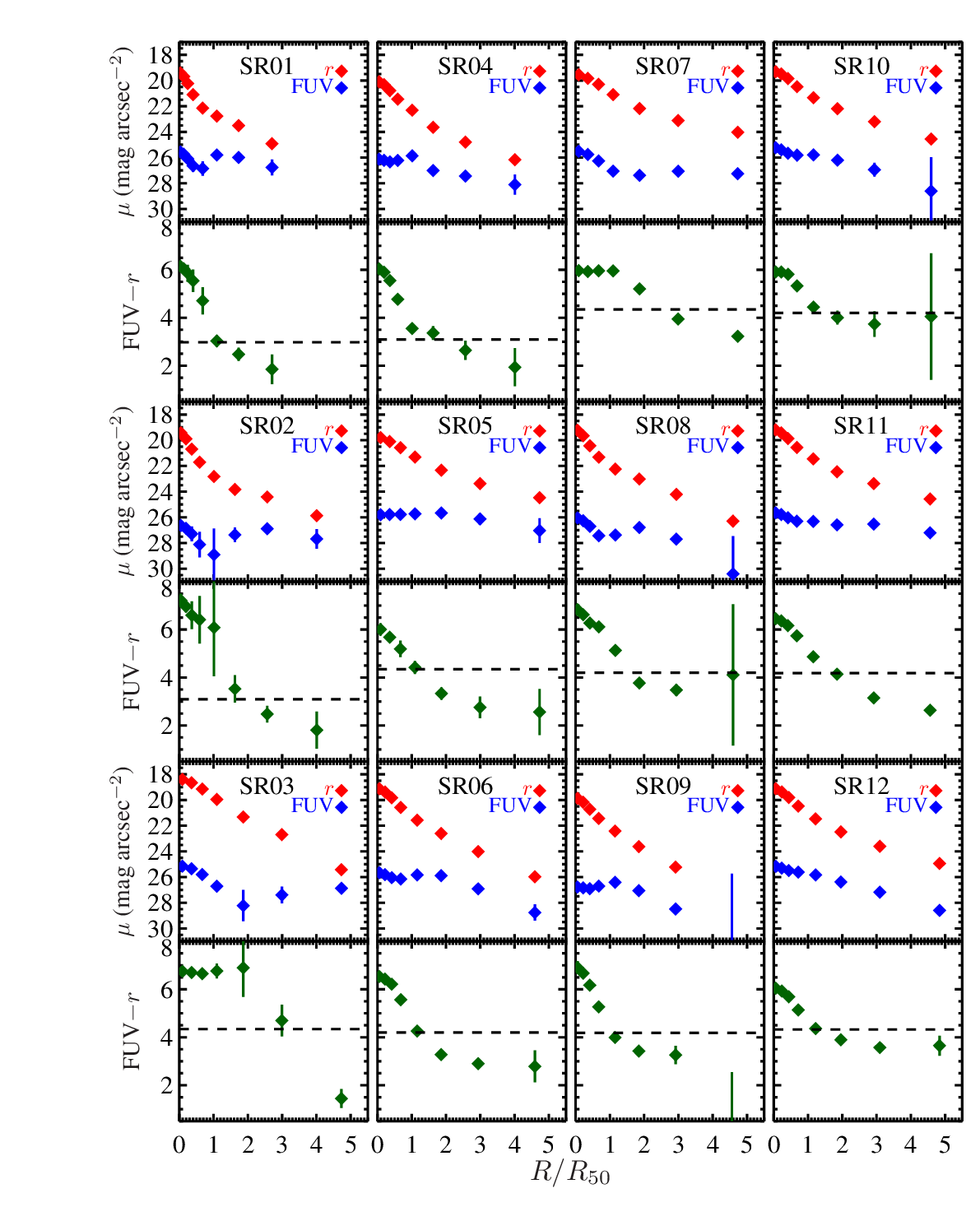}

  \caption{FUV and $r$-band surface brightness and \fuvr color profiles of the \esf. The $x$-axis is scaled to the half-light radius ($R_{50}$) of each galaxy. The dashed line indicates the integrated \fuvr color of each galaxy from Table \ref{table_prop}. The general shape of the color profiles is similar among the \esf; the centers are typically red (\fuvr$\ga6$), and the outer regions are quite blue ($2\la$ \fuvr$\la4$). Photometry has been corrected for Galactic extinction but not $k$-corrected. Measurements are tabulated in Table \ref{table_sb}.}
   
  \label{plot_prof}
\end{figure*}

\begin{figure*} 
  \epsscale{1.0}
  \figurenum{\ref{plot_prof}}
  \plotone{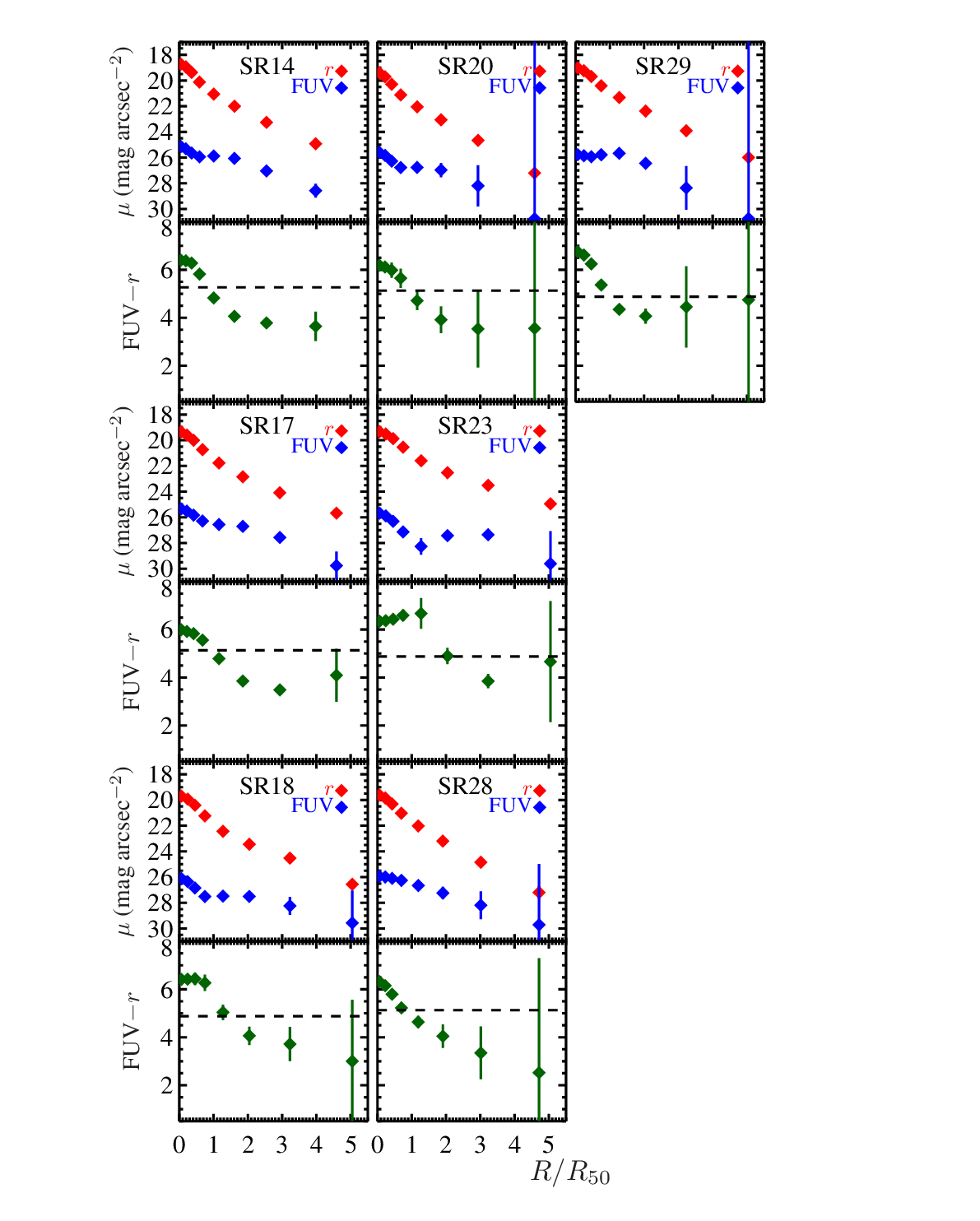}

  \caption{\emph{Continued}}
\end{figure*}

\begin{deluxetable*}{cccccccccc} 
\tablewidth{0in}
\tabletypesize{\scriptsize}

\tablecaption{Surface Brightness Measurements of the \esf \label{table_sb}}
\tablehead{ \colhead{Object} &  & \colhead{$R=0\farcs23$} 
			& \colhead{$R=0\farcs68$}
			& \colhead{$R=1\farcs03$}
			& \colhead{$R=1\farcs76$}
			& \colhead{$R=3\farcs00$}
			& \colhead{$R=4\farcs63$}
			& \colhead{$R=7\farcs43$}
			& \colhead{$R=11\farcs42$} }
							
\startdata
SR01  & $\mu_\mathrm{FUV}$ &  $ 26.040 \pm 0.193 $  &  $   26.221 \pm 0.160 $  &  $   26.574 \pm 0.353 $  &  $   27.120 \pm 0.468 $  &  $   27.339 \pm 0.568 $  &  $   26.283 \pm 0.249 $  &  $   26.470 \pm 0.277 $  &  $   27.253 \pm 0.611 $     \\
         & $\mu_r$ &  $ 19.571 \pm 0.027 $  &  $   19.854 \pm 0.006 $  &  $   20.393 \pm 0.015 $  &  $   21.256 \pm 0.014 $  &  $   22.308 \pm 0.015 $  &  $   22.928 \pm 0.026 $  &  $   23.668 \pm 0.027 $  &  $   25.078 \pm 0.090 $ \\
SR02  & $\mu_\mathrm{FUV}$ &  $ 27.000 \pm 0.288 $  &  $   27.214 \pm 0.242 $  &  $   27.653 \pm 0.577 $  &  $   28.481 \pm 0.990 $  &  $   29.267 \pm 2.028 $  &  $   27.721 \pm 0.566 $  &  $   27.248 \pm 0.343 $  &  $   28.043 \pm 0.766 $     \\
         & $\mu_r$ &  $ 19.542 \pm 0.026 $  &  $   20.010 \pm 0.005 $  &  $   20.814 \pm 0.017 $  &  $   21.828 \pm 0.018 $  &  $   22.944 \pm 0.032 $  &  $   23.953 \pm 0.032 $  &  $   24.534 \pm 0.045 $  &  $   25.995 \pm 0.114 $     \\
SR03  & $\mu_\mathrm{FUV}$ &  $ 25.276 \pm 0.118 $  &  $   25.484 \pm 0.079 $  &  $   25.927 \pm 0.176 $  &  $   26.844 \pm 0.310 $  &  $   28.348 \pm 1.214 $  &  $   27.519 \pm 0.654 $  &  $   27.002 \pm 0.380 $  &  $   27.979 \pm 1.004 $     \\
         & $\mu_r$ &  $ 18.442 \pm 0.030 $  &  $   18.697 \pm 0.004 $  &  $   19.188 \pm 0.008 $  &  $   19.994 \pm 0.018 $  &  $   21.363 \pm 0.050 $  &  $   22.736 \pm 0.068 $  &  $   25.469 \pm 0.114 $  &  $   29.288 \pm 0.616 $     \\
SR04  & $\mu_\mathrm{FUV}$ &  $ 26.358 \pm 0.192 $  &  $   26.446 \pm 0.113 $  &  $   26.570 \pm 0.186 $  &  $   26.454 \pm 0.125 $  &  $   26.092 \pm 0.087 $  &  $   27.241 \pm 0.285 $  &  $   27.668 \pm 0.394 $  &  $   28.336 \pm 0.781 $     \\
         & $\mu_r$ &  $ 20.169 \pm 0.037 $  &  $   20.395 \pm 0.002 $  &  $   20.863 \pm 0.010 $  &  $   21.529 \pm 0.021 $  &  $   22.386 \pm 0.029 $  &  $   23.718 \pm 0.046 $  &  $   24.871 \pm 0.070 $  &  $   26.243 \pm 0.138 $     \\
SR05  & $\mu_\mathrm{FUV}$ &  $ 26.303 \pm 0.251 $  &  $   26.289 \pm 0.210 $  &  $   26.278 \pm 0.342 $  &  $   26.234 \pm 0.268 $  &  $   26.169 \pm 0.252 $  &  $   26.637 \pm 0.448 $  &  $   27.536 \pm 0.960 $  &  $   29.510 \pm 6.347 $     \\
         & $\mu_r$ &  $ 19.960 \pm 0.035 $  &  $   20.269 \pm 0.003 $  &  $   20.746 \pm 0.009 $  &  $   21.475 \pm 0.016 $  &  $   22.499 \pm 0.037 $  &  $   23.547 \pm 0.040 $  &  $   24.635 \pm 0.075 $  &  $   26.979 \pm 0.361 $     \\
SR06  & $\mu_\mathrm{FUV}$ &  $ 25.985 \pm 0.154 $  &  $   26.134 \pm 0.071 $  &  $   26.369 \pm 0.107 $  &  $   26.476 \pm 0.067 $  &  $   26.158 \pm 0.042 $  &  $   26.213 \pm 0.047 $  &  $   27.244 \pm 0.110 $  &  $   29.085 \pm 0.631 $     \\
         & $\mu_r$ &  $ 19.235 \pm 0.023 $  &  $   19.489 \pm 0.003 $  &  $   19.939 \pm 0.008 $  &  $   20.692 \pm 0.011 $  &  $   21.681 \pm 0.029 $  &  $   22.718 \pm 0.062 $  &  $   24.126 \pm 0.097 $  &  $   26.081 \pm 0.203 $     \\
SR07  & $\mu_\mathrm{FUV}$ &  $ 25.812 \pm 0.144 $  &  $   26.059 \pm 0.073 $  &  $   26.560 \pm 0.140 $  &  $   27.356 \pm 0.170 $  &  $   27.682 \pm 0.202 $  &  $   27.363 \pm 0.170 $  &  $   27.557 \pm 0.188 $  &  $   28.903 \pm 0.696 $     \\
         & $\mu_r$ &  $ 19.647 \pm 0.029 $  &  $   19.930 \pm 0.002 $  &  $   20.400 \pm 0.007 $  &  $   21.198 \pm 0.008 $  &  $   22.279 \pm 0.015 $  &  $   23.215 \pm 0.026 $  &  $   24.130 \pm 0.037 $  &  $   26.025 \pm 0.099 $     \\
SR08  & $\mu_\mathrm{FUV}$ &  $ 26.269 \pm 0.176 $  &  $   26.486 \pm 0.088 $  &  $   26.926 \pm 0.163 $  &  $   27.651 \pm 0.191 $  &  $   27.593 \pm 0.162 $  &  $   27.007 \pm 0.107 $  &  $   27.912 \pm 0.227 $  &  $   30.621 \pm 2.943 $     \\
         & $\mu_r$ &  $ 19.291 \pm 0.023 $  &  $   19.721 \pm 0.004 $  &  $   20.507 \pm 0.013 $  &  $   21.391 \pm 0.014 $  &  $   22.319 \pm 0.016 $  &  $   23.092 \pm 0.030 $  &  $   24.289 \pm 0.062 $  &  $   26.368 \pm 0.141 $     \\
SR09  & $\mu_\mathrm{FUV}$ &  $ 27.119 \pm 0.259 $  &  $   27.186 \pm 0.118 $  &  $   27.244 \pm 0.159 $  &  $   27.043 \pm 0.083 $  &  $   26.751 \pm 0.055 $  &  $   27.397 \pm 0.106 $  &  $   28.830 \pm 0.358 $  &  $   31.768 \pm 5.695 $     \\
         & $\mu_r$ &  $ 19.979 \pm 0.036 $  &  $   20.287 \pm 0.006 $  &  $   20.845 \pm 0.009 $  &  $   21.548 \pm 0.019 $  &  $   22.531 \pm 0.028 $  &  $   23.741 \pm 0.067 $  &  $   25.340 \pm 0.134 $  &  $   35.000 \pm 1.918 $     \\
SR10  & $\mu_\mathrm{FUV}$ &  $ 25.440 \pm 0.137 $  &  $   25.606 \pm 0.099 $  &  $   25.876 \pm 0.196 $  &  $   26.025 \pm 0.177 $  &  $   25.997 \pm 0.170 $  &  $   26.414 \pm 0.287 $  &  $   27.154 \pm 0.532 $  &  $   28.814 \pm 2.633 $     \\
         & $\mu_r$ &  $ 19.388 \pm 0.025 $  &  $   19.557 \pm 0.003 $  &  $   19.924 \pm 0.006 $  &  $   20.554 \pm 0.009 $  &  $   21.411 \pm 0.010 $  &  $   22.270 \pm 0.022 $  &  $   23.278 \pm 0.027 $  &  $   24.625 \pm 0.072 $     \\
SR11  & $\mu_\mathrm{FUV}$ &  $ 25.870 \pm 0.144 $  &  $   26.006 \pm 0.065 $  &  $   26.253 \pm 0.096 $  &  $   26.522 \pm 0.064 $  &  $   26.537 \pm 0.050 $  &  $   26.809 \pm 0.067 $  &  $   26.742 \pm 0.057 $  &  $   27.425 \pm 0.112 $     \\
         & $\mu_r$ &  $ 19.269 \pm 0.023 $  &  $   19.500 \pm 0.002 $  &  $   19.943 \pm 0.006 $  &  $   20.637 \pm 0.011 $  &  $   21.525 \pm 0.028 $  &  $   22.525 \pm 0.028 $  &  $   23.447 \pm 0.026 $  &  $   24.642 \pm 0.056 $     \\
SR12  & $\mu_\mathrm{FUV}$ &  $ 25.368 \pm 0.137 $  &  $   25.501 \pm 0.060 $  &  $   25.695 \pm 0.083 $  &  $   25.818 \pm 0.046 $  &  $   26.035 \pm 0.038 $  &  $   26.578 \pm 0.060 $  &  $   27.372 \pm 0.105 $  &  $   28.789 \pm 0.403 $     \\
         & $\mu_r$ &  $ 19.189 \pm 0.022 $  &  $   19.438 \pm 0.001 $  &  $   19.873 \pm 0.005 $  &  $   20.546 \pm 0.010 $  &  $   21.537 \pm 0.024 $  &  $   22.551 \pm 0.042 $  &  $   23.669 \pm 0.057 $  &  $   25.004 \pm 0.104 $     \\
SR14  & $\mu_\mathrm{FUV}$ &  $ 25.856 \pm 0.144 $  &  $   26.042 \pm 0.066 $  &  $   26.373 \pm 0.103 $  &  $   26.667 \pm 0.068 $  &  $   26.604 \pm 0.048 $  &  $   26.788 \pm 0.058 $  &  $   27.764 \pm 0.124 $  &  $   29.298 \pm 0.537 $     \\
         & $\mu_r$ &  $ 18.968 \pm 0.019 $  &  $   19.185 \pm 0.003 $  &  $   19.607 \pm 0.008 $  &  $   20.359 \pm 0.012 $  &  $   21.294 \pm 0.026 $  &  $   22.243 \pm 0.070 $  &  $   23.494 \pm 0.141 $  &  $   25.169 \pm 0.284 $     \\
SR17  & $\mu_\mathrm{FUV}$ &  $ 25.763 \pm 0.137 $  &  $   25.934 \pm 0.062 $  &  $   26.255 \pm 0.095 $  &  $   26.707 \pm 0.069 $  &  $   26.983 \pm 0.061 $  &  $   27.122 \pm 0.070 $  &  $   27.989 \pm 0.137 $  &  $   30.186 \pm 1.094 $     \\
         & $\mu_r$ &  $ 19.489 \pm 0.026 $  &  $   19.733 \pm 0.005 $  &  $   20.144 \pm 0.008 $  &  $   20.869 \pm 0.007 $  &  $   21.913 \pm 0.010 $  &  $   22.987 \pm 0.015 $  &  $   24.225 \pm 0.038 $  &  $   25.813 \pm 0.125 $     \\
SR18  & $\mu_\mathrm{FUV}$ &  $ 26.289 \pm 0.183 $  &  $   26.535 \pm 0.113 $  &  $   27.029 \pm 0.253 $  &  $   27.688 \pm 0.339 $  &  $   27.656 \pm 0.319 $  &  $   27.685 \pm 0.377 $  &  $   28.419 \pm 0.693 $  &  $   29.746 \pm 2.523 $     \\
         & $\mu_r$ &  $ 19.758 \pm 0.029 $  &  $   19.993 \pm 0.005 $  &  $   20.472 \pm 0.016 $  &  $   21.299 \pm 0.024 $  &  $   22.497 \pm 0.042 $  &  $   23.504 \pm 0.066 $  &  $   24.580 \pm 0.150 $  &  $   26.619 \pm 0.397 $     \\
SR20  & $\mu_\mathrm{FUV}$ &  $ 25.800 \pm 0.160 $  &  $   26.036 \pm 0.133 $  &  $   26.483 \pm 0.317 $  &  $   26.979 \pm 0.399 $  &  $   26.974 \pm 0.395 $  &  $   27.183 \pm 0.553 $  &  $   28.407 \pm 1.601 $  &  $   30.964 \pm18.104 $     \\
         & $\mu_r$ &  $ 19.470 \pm 0.025 $  &  $   19.780 \pm 0.003 $  &  $   20.362 \pm 0.006 $  &  $   21.193 \pm 0.018 $  &  $   22.124 \pm 0.024 $  &  $   23.128 \pm 0.073 $  &  $   24.731 \pm 0.134 $  &  $   27.270 \pm 0.238 $     \\
SR23  & $\mu_\mathrm{FUV}$ &  $ 26.002 \pm 0.155 $  &  $   26.209 \pm 0.084 $  &  $   26.635 \pm 0.166 $  &  $   27.459 \pm 0.239 $  &  $   28.591 \pm 0.638 $  &  $   27.750 \pm 0.337 $  &  $   27.682 \pm 0.296 $  &  $   29.933 \pm 2.523 $     \\
         & $\mu_r$ &  $ 19.443 \pm 0.027 $  &  $   19.621 \pm 0.003 $  &  $   19.985 \pm 0.004 $  &  $   20.650 \pm 0.007 $  &  $   21.702 \pm 0.011 $  &  $   22.635 \pm 0.010 $  &  $   23.614 \pm 0.025 $  &  $   25.056 \pm 0.032 $     \\
SR28  & $\mu_\mathrm{FUV}$ &  $ 26.211 \pm 0.187 $  &  $   26.265 \pm 0.117 $  &  $   26.367 \pm 0.195 $  &  $   26.527 \pm 0.174 $  &  $   26.925 \pm 0.244 $  &  $   27.514 \pm 0.483 $  &  $   28.465 \pm 1.086 $  &  $   29.990 \pm 4.745 $     \\
         & $\mu_r$ &  $ 19.706 \pm 0.030 $  &  $   19.934 \pm 0.006 $  &  $   20.393 \pm 0.016 $  &  $   21.121 \pm 0.022 $  &  $   22.106 \pm 0.059 $  &  $   23.288 \pm 0.089 $  &  $   24.935 \pm 0.178 $  &  $   27.282 \pm 0.483 $     \\
SR29  & $\mu_\mathrm{FUV}$ &  $ 25.890 \pm 0.169 $  &  $   25.975 \pm 0.125 $  &  $   26.052 \pm 0.214 $  &  $   25.897 \pm 0.148 $  &  $   25.785 \pm 0.132 $  &  $   26.566 \pm 0.311 $  &  $   28.478 \pm 1.693 $  &  $   30.862 \pm16.334 $     \\
         & $\mu_r$ &  $ 19.057 \pm 0.020 $  &  $   19.283 \pm 0.004 $  &  $   19.731 \pm 0.007 $  &  $   20.452 \pm 0.012 $  &  $   21.360 \pm 0.020 $  &  $   22.417 \pm 0.034 $  &  $   23.946 \pm 0.077 $  &  $   26.034 \pm 0.110 $     

\enddata

\tablecomments{\emph{HST} FUV and SDSS $r$-band surface brightnesses (in mag arcsec$^{-2}$) are measured in circular annuli with outer radii specified in the column headings. Measurements are corrected for Galactic extinction but are not $k$-corrected. Negative fluxes are assigned the value $\mu=35.0$.}

\end{deluxetable*}

Another feature seen in Figure \ref{plot_prof} is the broad similarity in the overall shape of the color profiles of the \esf. The \fuvr color profiles show that their centers are quite red: they typically have a color FUV$-r\gtrsim6$, which is consistent with old, red sequence stellar populations \citep[Figure \ref{fuvr_mass};][]{rich05,donas07}. This is not too surprising given that the \esf\ were selected to have no central emission. By contrast, their outer parts ($R/R_{50}\ga2$, or $\ga7$ kpc) show very blue \fuvr colors, typically between 2 and 4. These are values typical of star-forming, blue cloud galaxies (Figure \ref{fuvr_mass}). Note that, despite the visual prominence of the central UV emission in the images, it only accounts for $\sim5\%$ of the total UV flux in each galaxy and negligibly contributes to the integrated \fuvr colors. A similar observation was made by \citet{kauffmann07}, who showed that the UV flux in their sample of GV ETGs was primarily due to SF on extended scales, even in objects with AGN activity.

\subsection{Stellar Population Analysis of the \esf} \label{sps_ext}

The color profiles in Figure \ref{plot_prof} quantitatively show that young stars exist in the outer disks of the \esf, confirming the result in \sr\ and Paper I. However, one cannot immediately determine whether those stars represent the vestiges of the original, fading SF or whether those young stars were produced in a rejuvenated burst of SF following a quiescent phase. Fortunately, a key question that \emph{can} be answered using broadband photometry is whether the outer disks consist purely of young stars or whether the young-star light is frosting on top of an older stellar population. 

\begin{figure} 
	\epsscale{1.2}
	\plotone{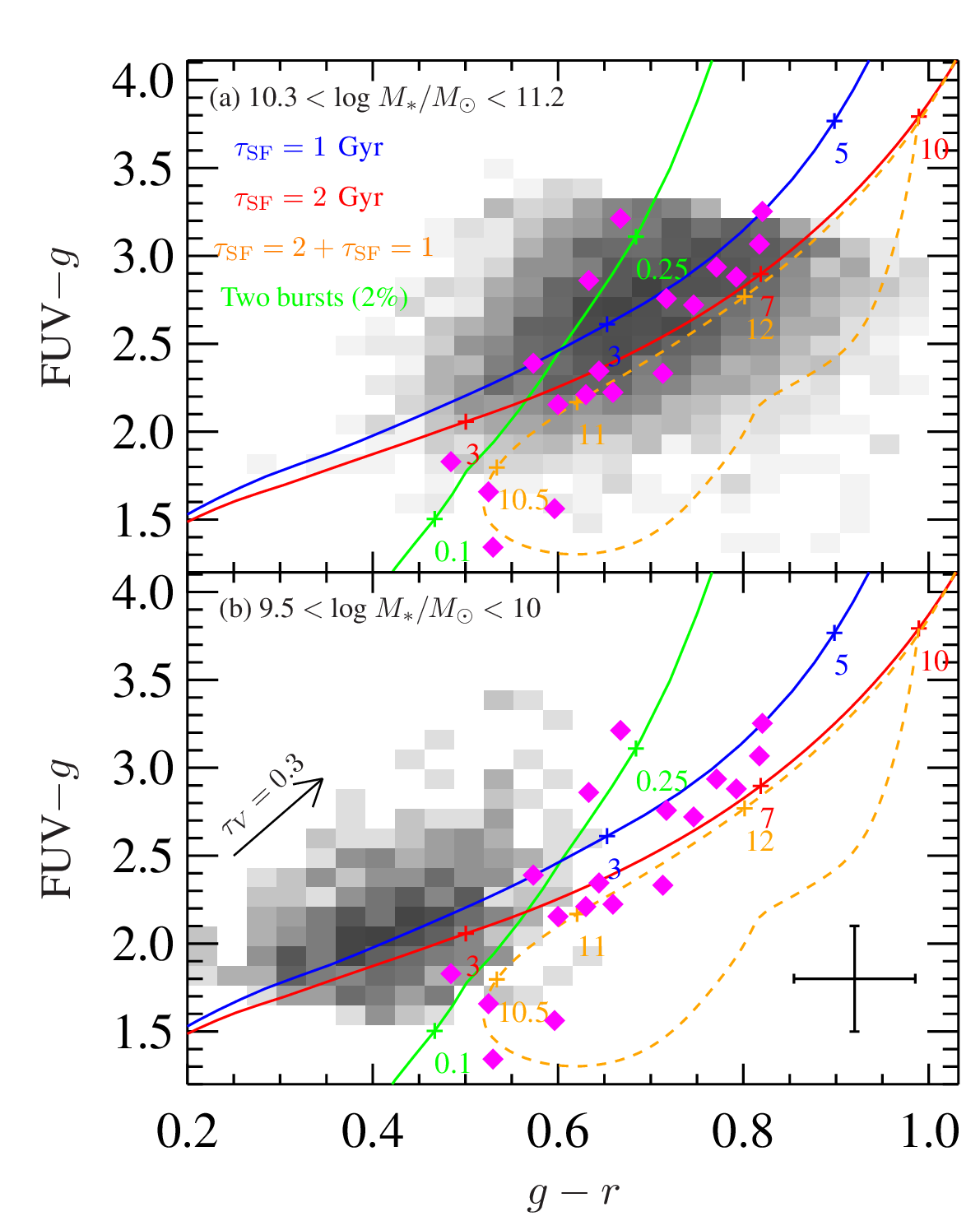}
	
	\caption{FUV$-g$ vs.~$g-r$ color-color diagrams. In both panels, the outer disk colors of the \esf\ are indicated (magenta diamonds), with median errors indicated by the cross in the bottom panel. Integrated colors of two SDSS comparison samples are shown: (a) high-mass and (b) low-mass blue cloud galaxies; these samples are indicated by the boxes in Figure \ref{fuvr_mass}. Various Bruzual \& Charlot models (age = 12 Gyr at $z=0.1$) are shown as colored lines, with ages indicated in Gyr (for the two-burst model, ages are after the second burst). The \citet{calzetti00} reddening vector is indicated for reference. The ESF-ETG outer colors are well-matched by the declining $\tau$-models (blue, red, and orange tracks) and also by the high-mass comparison sample, both of which contain old stars. The two-burst model (green track) also contains old stars, but the colors evolve very rapidly, implying that bursty rejuvenation is unlikely the cause of the SF in most \esf\ (Section \ref{sect_gv}). }

	\label{models} 
\end{figure}

Figure \ref{models} presents color-color diagrams showing the \emph{outer} colors of the \esf\ (magenta diamonds). Depending on the UV extent of each galaxy, the outer colors (FUV$-g$ and $g-r$) were computed by adding up the flux within the annuli that enclosed the outer UV rings/arms. The typical range of radii was between 4\farcs63 and 11\farcs42. Note that the outer colors have been corrected for Galactic extinction but have not been $k$-corrected. 

Included in Figure \ref{models} is a selection of stellar population models constructed using the 2007 version of the \citet{bc03} synthesis code. The models were generated assuming (1) a Chabrier IMF, (2) solar metallicity, and (3) a formation age of 12 Gyr. The various tracks were chosen to represent a range of possible SF histories for the \esf\ (discussed in Section \ref{discussion}). The blue and red tracks in the figure are two exponentially declining SF histories ($\tau$-models), with e-folding times $\tau_\mathrm{SF}=1$ and 2 Gyr, respectively. The orange track describes a ``double-$\tau$'' model: the first $\tau$-model has $\tau_\mathrm{SF}=2$ Gyr on top of which a second $\tau_\mathrm{SF}=1$ Gyr model is added 10 Gyr later (which produces 7\% of the total stellar mass). All three of these tracks include the two-component dust model of \citet{charlot00}. The default dust model parameters were chosen, i.e., the effective optical depth suffered by stars younger than $10^7$ yr is assumed to be $\hat{\tau}_V=1.0$, and the fraction of the attenuation due to the diffuse ISM is taken to be $\mu=0.3$. Finally, the green track assumes an instantaneous burst at $t=0$ and a second instantaneous burst 0.25 Gyr before the epoch of observation (taken to be $z=0.1$) that produces 2\% of the total stellar mass. This model is dust-free. For consistency with the ESF-ETG outer colors, which have not been $k$-corrected, all the tracks show observed-frame colors at $z=0.1$, the median redshift of the \esf. 

Before comparing to the data, we first discuss the behavior of the models themselves. The $\tau$-models represent gradually declining SF and are meant to represent the gradual decline of SF expected in massive blue galaxies \citep[e.g.,][]{noeske07}. As Figure \ref{models} shows, at a fixed age, a longer $\tau_\mathrm{SF}$ results in bluer colors. This is because the SF rate at a given age is higher in longer $\tau_\mathrm{SF}$ models, and thus the colors ``hover'' at bluer values for longer than a model with shorter $\tau_\mathrm{SF}$. Overall, the $\tau$-models span about 7 Gyr in age for the ESF-ETG color range plotted in Figure \ref{models}. The double-$\tau$ track is identical to the $\tau_\mathrm{SF}=2$ Gyr track up to $t=10$ Gyr, when the second episode of SF begins. The colors then become bluer before looping back and approaching the original $\tau_\mathrm{SF}=2$ Gyr track. In other words, the double-$\tau$ model is virtually indistinguishable from the single-$\tau$ model $\sim1$ Gyr after the second SF episode begins.

By contrast, the behavior of the two-burst model is rather different and simulates the effect of rapidly rejuvenated SF in a quiescent galaxy. The second burst occurs on top of an evolved stellar population, and thus affects the FUV$-g$ color (which is sensitive to young stars) more strongly than the $g-r$ color (which is dominated by older stars). This causes the track to rapidly redden in FUV$-g$ as the OB stars die off while the optical $g-r$ color changes little. As seen in the figure, the FUV$-g$ color reddens within $\sim200$ Myr, implying that any significant bluing in the UV due to bursty SF would have to be short-lived.

What do the models tell us about the stellar populations in the outer disks of the \esf? As seen in Figure \ref{models}(a), the outer ESF-ETG colors are generally consistent with the declining $\tau$-models (both single-$\tau$ and double-$\tau$) and imply intermediate to old ages for the underlying stellar populations in these galaxies. By contrast, the two-burst model colors are somewhat offset from the mean $g-r$ colors of the \esf. One could argue that adjusting the mass fraction of the second burst (set to 2\% in the figure) or incorporating dust reddening could bring the colors more in agreement. However, regardless of the values used, bursts evolve \emph{very quickly} through the region occupied by the \esf, and thus it is statistically unlikely that most \esf\ are bursts unless they are exceedingly rare objects. We return to this issue in Section \ref{sect_gv} and show that this is probably not the case. 

Collectively, this comparison shows that the SF in the outer disks of the \esf\ is taking place \emph{in regions with old stars}. Moreover, it demonstrates that \emph{bursty}  rejuvenation is not likely responsible for the SF we see. This last point is consistent with SR2010's selection against galaxies with post-starburst signatures or optical disturbances indicative of recent, short-duration bursts.

In addition to the model comparison above, it is useful to demonstrate the existence of old stellar populations in the ESF-ETG disks in an empirical manner. To that end, we now compare the outer disk colors of the \esf\ to two sets of galaxies chosen from the SDSS master sample. These sets are indicated by the rectangles in Figure \ref{fuvr_mass}. The comparison sample in Figure \ref{models}(a) consists of high-mass ($10.3<\log M_*/M_\odot<11.2$) galaxies matched in redshift to the \esf\ ($0.08<z<0.12$) that have the same \emph{integrated} \fuvr color as the \emph{outer} parts of the \esf, namely $2<$ \fuvr$<4$, i.e., they reside in the blue cloud. The integrated colors of these galaxies have been corrected for Galactic extinction but are not $k$-corrected. We note that the $k$-correction to $z=0.1$ is small, i.e., $<0.1$ mag for FUV$-g$ and $<0.04$ mag for $g-r$. As can be seen, the outer ESF-ETG colors are fully consistent with the integrated colors of these comparison galaxies. Because the latter are massive, they have significant old stellar populations, and hence red $g-r$ colors (the effect of dust attenuation is discussed below). This bolsters the conclusion from comparing to models above that old stars are present in the outer disks of the \esf. 

By contrast, the sample in Figure \ref{models}(b) is comprised of low-mass ($9.5<\log M_*/M_\odot <10.0$) blue cloud galaxies within the same redshift range as the \esf\ ($0.08<z<0.12$). Their $g-r$ colors seem systematically too blue compared to the \esf, reflecting the fact that they have higher specific SF rates than their more massive counterparts \citep[e.g.,][]{salim07}. This results in a higher fraction of younger, and hence bluer, stars in the low-mass objects, driving their $g-r$ color blueward. Indeed, visual inspection of the SDSS color images of the low- and high-mass comparison galaxies clearly shows that the former have blue disks, while the latter have red bulges and disks. The point is that the integrated colors of more massive, optically redder comparison galaxies are a good match to the outer colors of the \esf\ whereas the less massive, optically bluer comparison sample does not match. Thus we have seen that both data and theory lead to the same conclusion: old stars are abundant in the outer star-forming regions of the \esf.

Our discussion so far has neglected the effects of dust attenuation on the observed colors of the comparison galaxies. Indeed, the amount of dust attenuation in SF galaxies increases with stellar mass \citep[e.g.,][]{brinchmann04,salim07,wuyts11}, potentially complicating our comparison with the \esf, especially for the massive sample shown in Figure \ref{models}(a). We attempted to remove the most dust-reddened objects by excluding edge-on systems with axis ratio $b/a<0.5$. This makes no difference in the distribution of points in Figure \ref{models}. In any case, assuming that the dust-corrected colors of the comparison galaxies are bluer than shown in Figure \ref{models} only strengthens our conclusion that the outer disks of the \esf\ contain \emph{old} stars. 

\subsection{Stacked SDSS Spectra} \label{sect_spec}

The SDSS fiber spectra, which probe the central 3\arcsec\ ($\sim5$ kpc at $z\sim0.1$), can provide additional insight into the stellar populations and any potential emission activity that may be responsible for the central UV light seen in nearly all the \esf. Also, they can inform our discussion about how SF is evolving in the \esf, particularly the possible role of AGN feedback in shutting down SF \citep[e.g.,][]{schawinski07}. \sr\ tentatively ascribed the central UV emission to weak AGN activity based on a stacked BPT classification. Stacking was necessary because the \esf\ do not have securely detected emission lines, meaning any (potentially weak) emission features will not be readily apparent in the individual spectra. 

\begin{figure} 
	\epsscale{1.23}
	\plotone{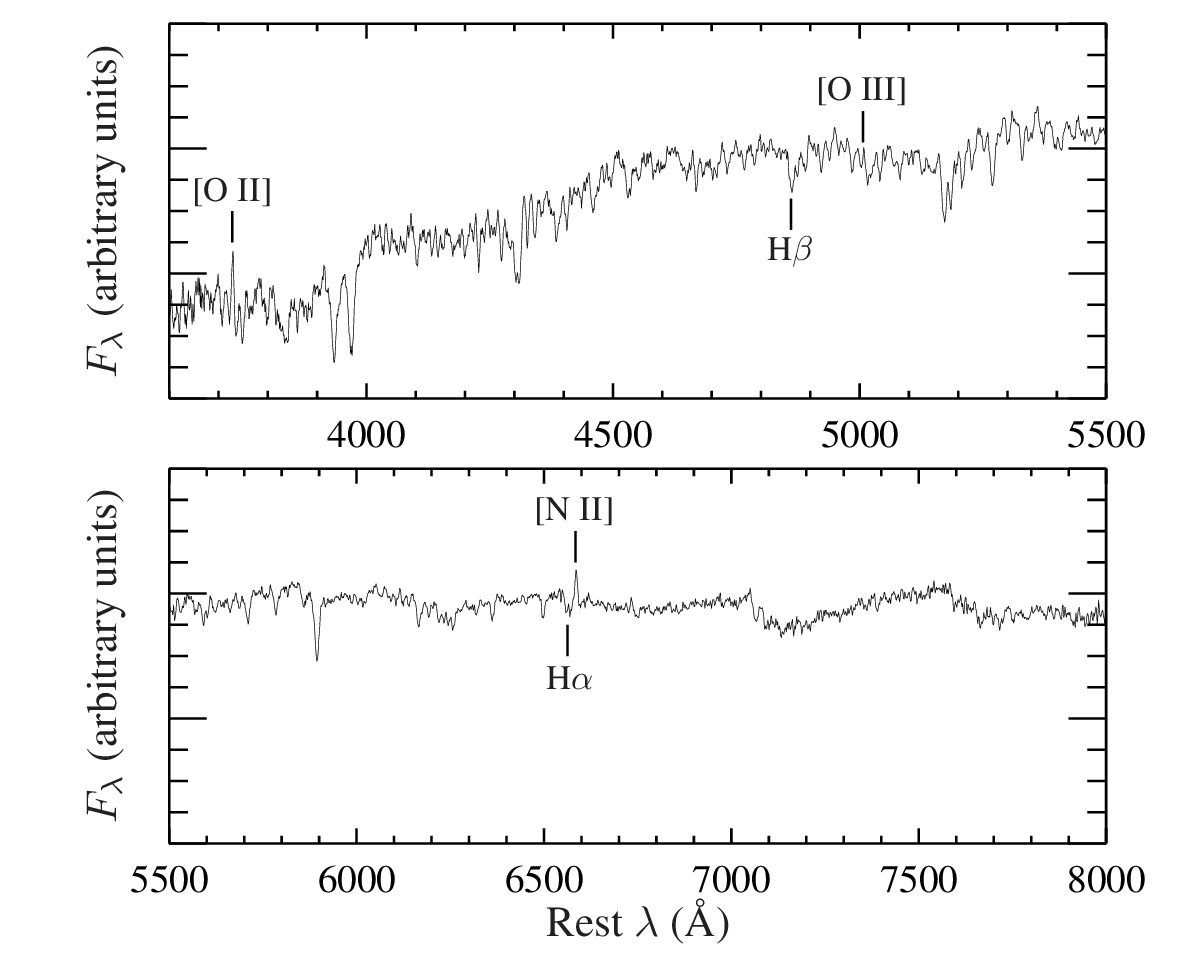}

	\caption{Stacked SDSS spectrum of the \esf. While the continuum is that due to an old stellar population, emission lines of [\ion{O}{2}] $\lambda\, 3727$ and [\ion{N}{2}] $\lambda\,6584$ are clearly detected, and [\ion{O}{3}] $\lambda\, 5007$ is barely visible. These features are characteristic of LINER emission. H$\alpha$ and H$\beta$ are also labeled for reference. The strength of the H$\beta$ absorption line, together with $\langle$Fe$\rangle$, is used to estimate an average luminosity-weighted stellar population age of $\sim 7$ Gyr (see text for details). }
	
	\label{ext_spec}
\end{figure}

We reach similar conclusions based on stacking the spectra themselves. The stacking method is detailed in \citet{graves09}. Figure \ref{ext_spec} shows the resulting stacked spectrum of the \esf. Interestingly, weak emission lines due to [\ion{O}{2}] $\lambda\, 3727$ and [\ion{N}{2}] $\lambda\, 6584$ that were not apparent in the individual spectra are clearly evident in the stack. These emission lines are characteristic of LINER emission \citep[e.g.,][]{heckman80,yan06,graves07}. If LINERs are indeed a weak phase in the AGN life cycle, this might indicate that AGN feedback played a role in shutting down SF, at least in the centers of these galaxies. However, AGN feedback has historically been proposed as a mechanism that quenches SF \emph{globally} in a galaxy, not just in the center \citep[e.g.,][]{dimatteo05,hopkins06}. Given the extended SF seen in the outer parts of the \esf, it would appear that AGN feedback is most effective at removing (or heating up) the \emph{central} gas and less so at disrupting SF in the outer parts. Interestingly, recent observations of X-ray-selected AGN hosts at $z\sim1$ may suggest that AGN feedback quenches SF only within the central kpc \citep{ammons11}.  

It is also possible that the LINER emission observed in \esf\ is unrelated to AGN \citep[e.g.,][]{filippenko03}. It has been suggested that LINER emission is dominated by emission from hot gas in the ISM ionized by extended UV sources, such as post-AGB stars, rather than an AGN \citep[e.g.,][]{binette94,goudfrooij94,sarzi10,yan11}. In any case, the emission in the stacked spectrum is both weak and uncharacteristic of young stars, which is the point we wish to establish.

With our resolved photometry we are able to say more about the nature of the central UV emission in the \esf. The central \fuvr colors of the \esf\ (\fuvr$\ga6$) in Figure \ref{plot_prof} are consistent with old, red sequence galaxies \citep[Figure \ref{fuvr_mass};][]{rich05,donas07}. The possibility that LINER emission is due to ionization by old, UV-bright (e.g., post-AGB) stars bolsters the observation that the central UV emission is not due to recent SF. To summarize, the stacked spectra indicate weak LINER-like activity, but the red \fuvr colors do not require anything besides old UV upturn stars.

With spectra one can also compute mean ages of the central stellar populations from absorption features. To gain quantitative insight into the typical age of the stars in the central regions of the \esf, the publicly available code \texttt{EZ\_Ages} \citep{schiavon07,graves08} was used to compute a mean luminosity-weighted age from the stacked spectrum. Stellar absorption templates are fit to the spectrum, and an age is derived based on the calculated strength of age-sensitive absorption features, such as H$\beta$. Caution must be taken to account for any absorption line infill due to nearby emission. The H$\beta$ absorption line strength measured here was corrected for infill by first determining the EW of the [\ion{O}{3}] emission [visible only in the continuum-subtracted spectrum (not shown)], then calculating the appropriate H$\beta$ correction using Equation (A1) of Appendix A in \citet{graves10}. The correction was subtracted from the H$\beta$ line strengths before estimating ages. The resulting mean luminosity-weighted age (determined using primarily H$\beta$ and $\langle$Fe$\rangle$) of the central regions of the \esf\ is $6.8^{+2.1}_{-1.7}$~Gyr. This derived age is younger than a pure, maximally old passively evolving stellar population, but it is consistent with light-weighted ages of the central regions of other ETGs in the local universe \citep[e.g.,][]{trager00}. Moreover, an age of $\sim7$ Gyr is not consistent with any significant contribution from young stars. For example, the SAURON galaxies with recently formed young stellar populations have H$\beta$-based light-weighted ages younger than $\sim3$ Gyr \citep{kuntschner10}. 

\subsection{The SR2010 \esf\ in Context} \label{sect_gv}

With their characteristic blue outer colors and distinctive UV rings, the \esf\ represent a clearly identifiable stage of evolution of massive ETGs through the GV. While the UV properties of only a small number of \esf\ have been presented in this work, it is useful, if possible, to place our results in the broader context of evolution through the GV. To do so requires answering two questions. First, are \esf\ rare or common among the GV population? Second, what is the ``residence time'' of \esf\ in the GV? The answer to the first question can help establish whether or not extended SF is a dominant phenomenon among GV galaxies. Answering the second question can help distinguish between plausible SF histories of the \esf, such as those discussed in Section \ref{sps_ext}.

Accordingly, in this section a search is conducted for further potential ESF-ETG candidates in the SDSS master sample defined in Section \ref{sect_ancillary}. We first review the selection cuts made by \sr\ in defining their \hst sample and explain the motivations behind them. Then we apply a modified version of the \sr\ cuts on the SDSS master sample and examine the properties of the resulting objects, many hundreds of which we also believe to be \esf. Sources of contamination that need to be accounted for when counting GV galaxies are discussed, and a rough estimate of the fraction of \esf\ in the GV is presented. This is followed by a discussion of the ESF-ETG GV residence time and its use in constraining potential SF histories for the \esf.

\subsubsection{\sr's Selection Criteria}

At first blush, the small number of galaxies selected by \sr\ would seem to suggest that \esf\ comprise an insignificant part of the GV. In the following sections we show that this is not the case. Recall that \sr\ selected an initial sample of $\sim60$ objects with strong UV excess and no spectral signatures of SF or AGN activity starting with a cross-matched SDSS DR4/GALEX IR1.1 catalog \citep{seibert05} containing 18,127 galaxies\footnote{This number includes the subset of galaxies with both NUV and FUV detections and redshift $z<0.12$ located within 0\fdg55 of the center of the GALEX field of view. The catalog sky coverage is 645 deg$^2$.}. The second column of Table \ref{counts} shows the number of galaxies we obtain from the same catalog that satisfy each successive \sr\ cut, until the final sample is obtained. To review, the first cut made by \sr\ selected for ETGs by choosing galaxies with $r$-band concentration $C>2.5$. To exclude objects with emission due to SF or AGN, only galaxies in the ``Unclassifiable'' category as defined by \citet{brinchmann04} were kept (these are galaxies with generally weak emission). To further prune remaining galaxies with weak emission, \sr\ required H$\beta$ S/N $<3$ and excluded any galaxies located within the LINER region of the BPT diagram, i.e., $-0.2<\log$\,([\ion{N}{2}]/H$\alpha$) $< 0.5$ and $-0.3<\log\,$([\ion{O}{3}]/H$\beta$) $<0.8$, regardless of line S/N. Finally, a color cut was made to select galaxies with strong UV excess (\fuvr$<5.3$). As can be seen in the second column of Table \ref{counts}, their very small final sample is due mainly to the cuts on ``Unclassifiable'' galaxies and \fuvr color. We note that among these $\sim60$ objects are galaxies with clear optical disturbances, late-type morphologies and post-starburst signatures. To arrive at their final sample of 30 objects, SR2010 excluded these contaminants. This is discussed further in Section \ref{contamination}.

\begin{deluxetable}{lcc} 
\tablewidth{0in}
\tablecaption{Sample Sizes Using Original \sr\ ESF-ETG Selection Criteria \label{counts}}

\tablehead{ \colhead{Cut} & \colhead{DR4/IR1.1\tablenotemark{a} } &
			\colhead{SDSS Master Sample\tablenotemark{b}} }
			
\startdata
Initial Sample\tablenotemark{c} & 18127 (100\%)& 31763 (100\%)\\
Conc $>2.5$ & 8119 (44.8\%) &  15617 (49.2\%)  \\
``Unclassifiable''\tablenotemark{d} & 1819 (10.0\%) & 4152 (13.1\%)  \\
H$\beta$ S/N $<3$\tablenotemark{e} & 1761 (9.7\%) &  4026 (12.7\%) \\
Not LINER\tablenotemark{f} & 1008 (5.6\%) &  2338 (7.4\%)  \\
\fuvr$<5.3$ & \textbf{67} (0.37\%) &  \textbf{186} (0.59\%) 

\enddata

\tablenotetext{a}{SDSS/GALEX catalog from which \sr\ selected \esf.}
\tablenotetext{b}{SDSS DR7/GALEX GR6 catalog used in this work.}
\tablenotetext{c}{Redshift $0.005<z<0.12$, NUV- and FUV-detected in MIS, and within 0\fdg55 of the center of the GALEX field of view.}
\tablenotetext{d}{Galaxies lacking statistically significant emission lines according to \citet{brinchmann04}.}
\tablenotetext{e}{A further cut to ensure weak emission; see text.}
\tablenotetext{f}{``Unclassifiable'' galaxies outside the LINER region of the BPT diagram.}

\end{deluxetable}

We applied the \sr\ cuts on the SDSS master sample, which is based on SDSS DR7 and GALEX GR6 data and is roughly twice as large as the \citet{seibert05} dataset used by \sr. This results in a sample of 186 galaxies, as shown in the third column of Table \ref{counts}. Both samples include only galaxies with redshift $z<0.12$ that are detected in FUV at GALEX MIS depths. Starting with this larger sample yields a bigger bottom line, but apart from that the effect of the various cuts is very similar in the two samples. In other words, it appears that \esf\ selected using the strict \sr\ criteria are still rather uncommon, comprising only $\sim0.6\%$ of the entire SDSS master sample. It should be noted that the original goal of SR2010's selection strategy was to study objects with unusually strong UV excess that was not obviously related to SF or AGN activity. Therefore, they purposely imposed very strict criteria on spectroscopic quiescence to reduce possible UV contamination from SF or AGN. Their selection criteria were not designed to obtain a \emph{complete} sample of ESF-ETGs. To achieve that we modify some of their criteria below.

\subsubsection{Searching for Additional \esf\ in the Green Valley}

The hallmark of \esf\ is that they are characterized by extended SF in their outer parts, while their centers are devoid of SF. Nevertheless, as demonstrated in Section \ref{sect_spec}, their central regions may harbor weak emission consistent with LINERs. Regardless of the source of the LINER emission (Section \ref{sect_spec}), such emission may be a common feature of \esf\ and may hold clues to their evolution through the GV. Thus a search for \esf\ ought not to exclude \emph{a priori} galaxies with weak (LINER) emission since such emission unlikely affects the UV flux (Section \ref{sect_spec}). Because \sr\ explicitly excluded galaxies with \emph{any} central emission, their exclusion of LINERs probably also excluded many genuine non-star-forming galaxies. In what follows, we relax the \sr\ selection criteria to accept LINERs, and thereby demonstrate that a larger sample of ESF-candidates exists and that such a sample indeed shares many similarities with the \esf\ observed by \emph{HST}.

In selecting our new ``ESF-candidate sample'', the most salient \sr\ selection criteria were retained, namely, the morphological cut (concentration $C>2.5$) and the \fuvr color cut (\fuvr$<5.3$). However, we avoided selecting galaxies based on the spectral categories defined by \citet{brinchmann04}, which relied on S/N cuts and BPT classification. Instead, weak central-star-forming galaxies were chosen based on \halpha equivalent width (EW). Because the EW measures the continuum-normalized line strength, it can be used to compare the emission properties in galaxies with different intrinsic luminosities in a relatively uniform manner. Since we are interested in selecting (bright) galaxies with weak \halpha emission, the EW provides a cleaner indicator of \halpha line strength than a S/N cut on the emission line flux. 

\begin{figure} 
	\epsscale{1.2}
	\plotone{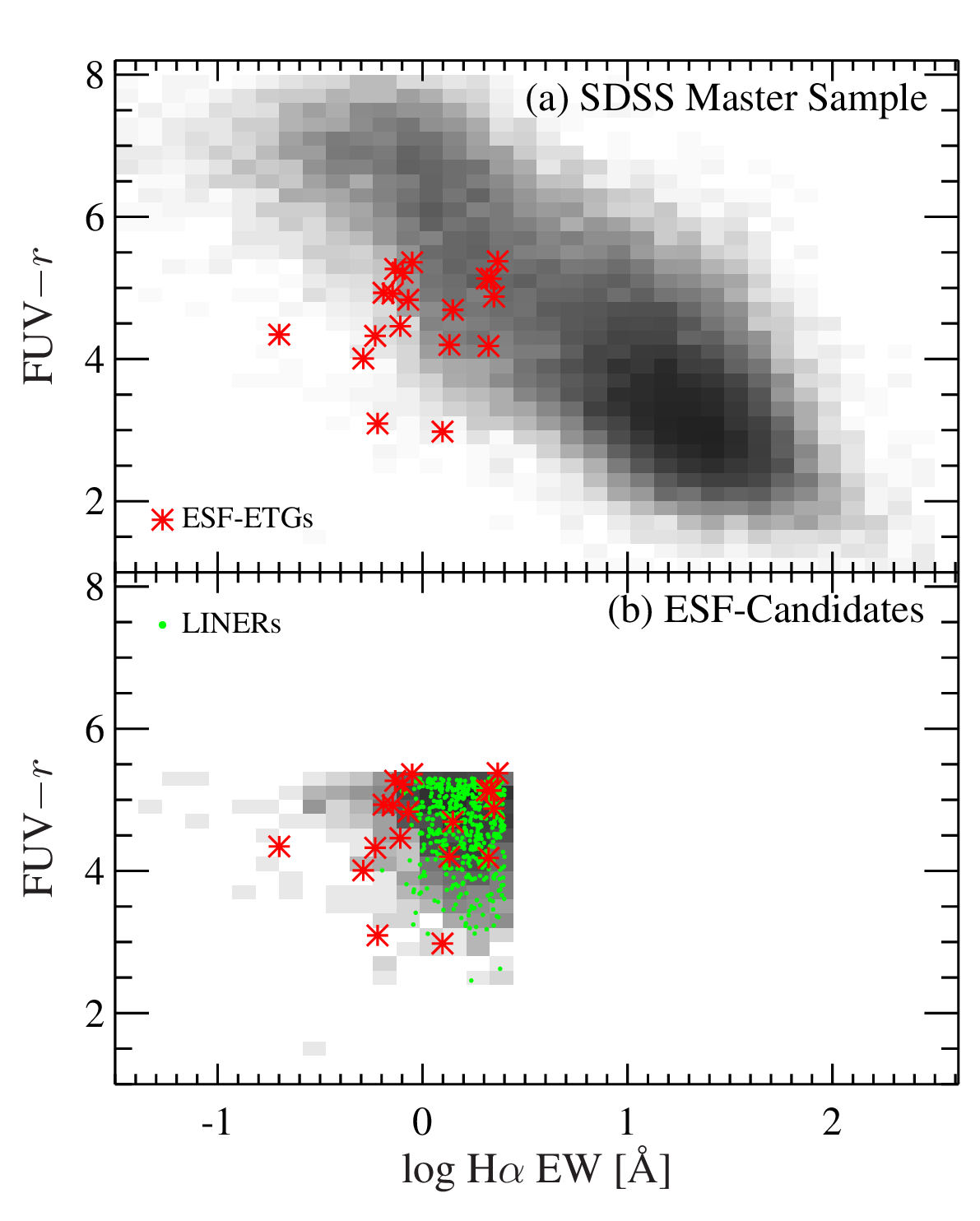}
	
	\caption{\fuvr vs.~\halpha equivalent width (EW). (a) The SDSS master sample (gray histogram) is plotted along with the \esf\ (red points). By selection, the \esf\ have \fuvr$<5.3$, concentration $C>2.5$, and weak \halpha emission. (b) A new ESF-candidate sample (gray histogram, $\sim1200$ objects) chosen to match the \esf\ and selected using \fuvr$<5.3$, $C>2.5$, and a simple cut on \halpha EW (log \halpha EW $<0.4$). Galaxies classified as LINERs by \citet{brinchmann04} (534 objects) are plotted as green dots. The distribution of the \esf\ in this parameter space is consistent with that of the ESF-candidates. Also, the LINERs fall generally within the same region as both the \esf\ and the ESF-candidate sample.} 
	
	\label{haew_fuvr}
\end{figure}

Shown in Figure \ref{haew_fuvr}(a) is the FUV$-r$ vs.~\halpha EW distribution of the SDSS master sample, with the \esf\ overplotted. We see that the \esf\ lie within (or below) the GV in \fuvr (by selection) and also have low \halpha EWs for their \fuvr color. The low EWs are consistent with \sr's selection against detectable emission lines using S/N cuts. The maximum \halpha EW among the \esf\ is approximately log \halpha EW [\AA] $\approx0.4$, and we adopt this as the upper boundary for the ESF-candidates. Thus our final sample of new ESF-candidate galaxies satisfies the following criteria: (1) \fuvr$<5.3$, (2) $i$-band concentration\footnote{SR2010 selected galaxies using $r$-band concentration. However, the difference between the $r$- and $i$-band concentrations is negligible.} $C>2.5$, and (3) log \halpha EW $<0.4$. Because galaxies are now included regardless of spectral classification, the ESF-candidate sample includes objects spectroscopically categorized as LINERs but still excludes star-forming galaxies, which have higher \halpha EWs. By casting a wider net, we now find that the ESF-candidate sample contains 1226 galaxies, almost seven times more than the \sr\ criteria returned. Moreover, like the \esf, the ESF-candidate galaxies lie on the optical red sequence despite their blue \fuvr colors. Table \ref{mycounts} lists the number of galaxies remaining after each of the new selection cuts. Evidently including galaxies with weak emission yields many more galaxies.

\begin{deluxetable}{lc} 
\tablewidth{0in}
\tablecaption{Sample Size Using Modified Criteria to Find ESF-Candidates \label{mycounts}}

\tablehead{\colhead{Cut} & \colhead{SDSS Master Sample}}

\startdata
Initial Sample & 31763 (100\%) \\
Conc $>2.5$ & 15617 (49.2\%) \\
log \halpha EW $<0.4$ & 6732 (21.2\%) \\
\fuvr$<5.3$ & \textbf{1226}\tablenotemark{a} (3.9\%)

\enddata

\tablenotetext{a}{Before correction for potential contamination, which reduces this number by half (Section \ref{contamination}).}

\end{deluxetable}

Figure \ref{haew_fuvr}(b) presents the distribution of the original \esf\ and the new ESF-candidate sample in \fuvr vs.~\halpha EW. ESF-candidates classified as LINERs according to \citet{brinchmann04} are plotted as green dots. Three things to note are: (1) The new ESF-candidate sample has a similar distribution as the \esf, (2) the LINERs make up a significant fraction of the ESF-candidate sample ($534/1226\approx44\%$), and (3) most of the original \esf\ have \halpha EWs comparable to the LINERs. Point (1) is simply due to selecting galaxies with similar color and EWs as the \esf. Point (2) suggests that many of the ESF-candidates have weak emission that in the stronger cases has been classified as LINERs. Point (3) is consistent with the results of Section \ref{sect_spec}, where LINER emission lines, including H$\alpha$, were detected in the stacked spectrum of the \esf. If one assumes that LINER activity is common in aging stellar populations, then it is reassuring to find that a significant fraction of the ESF-candidate sample are LINERs. Moreover, it is of interest to see whether the LINER galaxies have other properties in common with the \esf.

\begin{figure} 
	\epsscale{1.2}
	\plotone{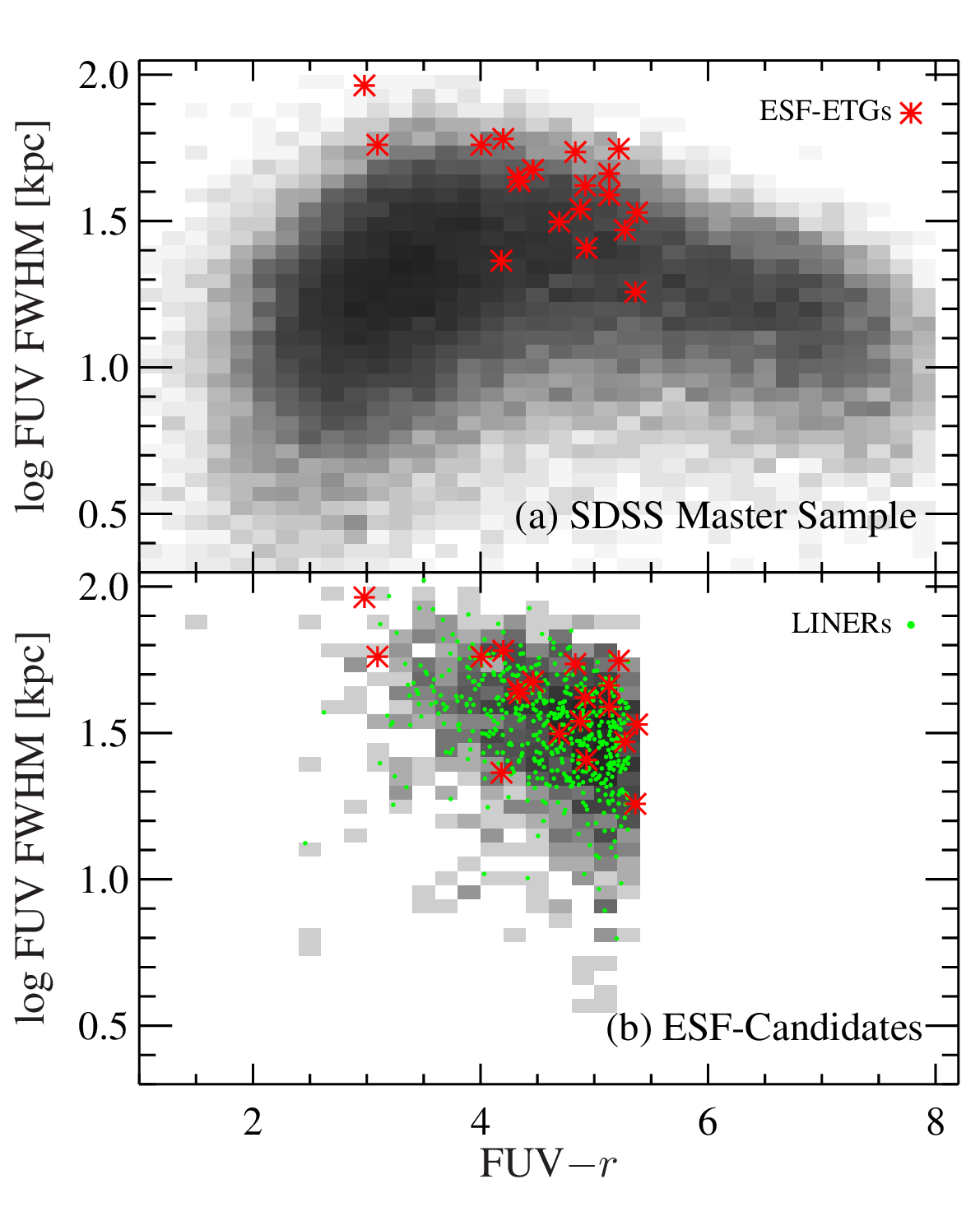}
	
	\caption{GALEX FUV FWHM (a measure of UV diameter) vs.~FUV$-r$. (a) The distribution of the entire SDSS master sample (gray histogram) and the \esf\ (red points). The finding of \sr\ is confirmed that the \esf\ tend to lie high when compared to all GV galaxies of similar color: their median size is $\sim0.38$ dex above the SDSS master sample. (b) The same distribution but now restricted to the ESF-candidate sample of $\sim1200$ galaxies chosen to match the original \esf\ with \fuvr$<5.3$, concentration $C>2.5$, and log \halpha EW $<0.4$ (gray histogram). Overplotted are galaxies identified as LINERs by \citet{brinchmann04} (green dots). The median offset in size of the \esf\ is now only 0.13 dex above a linear fit to the ESF-candidates.}
	
	\label{fuvr_fuvsize}
\end{figure}

A notable characteristic of the \esf\ is their large UV sizes (\sr). This is important because it may be a clue to the origin of the gas fueling the extended SF. \sr\ showed that the GALEX FUV FWHM sizes of the \esf\ (roughly their UV diameters) tend to be larger than other GV galaxies with similar specific SF rates (their Figure 5). We obtain a similar result in Figure \ref{fuvr_fuvsize}(a), where the distribution of FUV FWHM vs.~\fuvr color is plotted in a similar style as Figure \ref{haew_fuvr}. Figure \ref{fuvr_fuvsize}(a) shows the \esf\ compared to the full SDSS master sample. Clearly, the \esf\ are systematically larger than the average of \emph{all other GV galaxies of similar \fuvr color}. 

Since we aim to establish the similarity between the true \esf\ and the ESF-candidates, we proceed to compare the UV sizes of the \esf\ with the ESF-candidates. We refrain from making comparisons with the rest of the GV. Figure \ref{fuvr_fuvsize}(b) shows the UV sizes of the \esf\ and ESF-candidates. Since the latter were chosen in a similar way as the \esf, it is not too surprising that their UV sizes are roughly consistent with those of the \esf. The higher-\halpha galaxies in the GV prove to have smaller UV sizes, and agreement with the ESF-ETG sample is improved when they are eliminated.

To quantify any remaining offset between the ESF-candidates and the \esf, a linear fit was made to the ESF-candidate sample and the residuals in FUV FWHM of the \esf\ were computed. The median FUV FWHM residual for the \esf\ is $0.13\pm0.03$ dex. This implies that the \esf\ are $\sim35\%$ larger than the ESF-candidate galaxies. This slightly larger size may be because the \esf\ were chosen to satisfy strict criteria that may have biased the sample to larger UV sizes, whereas the ESF-candidate sample has not been vetted in a similarly stringent manner. In particular, SR2010 excluded post-starburst galaxies and other blue interlopers, which tend to have smaller UV sizes. We also note that the LINERs in Figure \ref{fuvr_fuvsize}(b) are not systematically larger or smaller than the ESF-candidates or the \esf. This similarity between the \esf\ and LINERs reinforces our assertion that many of the LINER galaxies have large UV diameters consistent with the \esf, and that a larger population of GV galaxies with extended UV emission exists.

\begin{figure} 
	\epsscale{1.2}
	\plotone{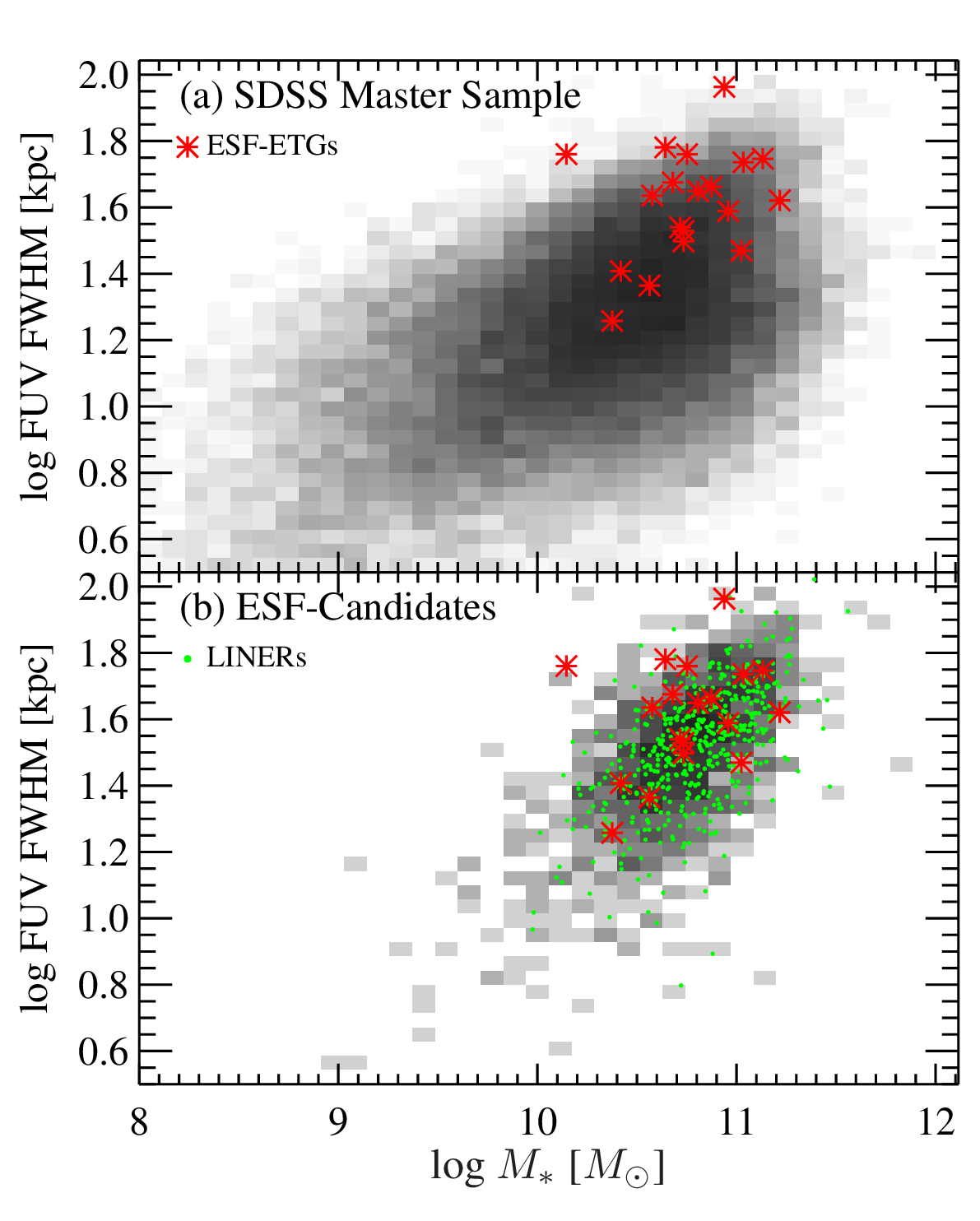}
	
	\caption{Galaxy stellar mass $M_*$ vs.~GALEX FUV FWHM (a measure of UV diameter). (a) The entire SDSS master sample (gray histogram) and the \esf\ (red points). The \esf\ lie high by $\sim0.3$ dex. (b) The same distribution but now restricted to the ESF-candidate sample of $\sim1200$ galaxies chosen to match the original \esf\ with \fuvr$<5.3$, concentration $C>2.5$, and log \halpha EW $<0.4$ (gray histogram). Overplotted are galaxies identified as LINERs by \citet{brinchmann04} (green dots). The median offset in UV size of the \esf\ is now only 0.08 dex above a linear fit to the ESF-candidates. } 
	
	\label{mass_fuvsize}
\end{figure}

An additional comparison is made in Figure \ref{mass_fuvsize}, where we show the distribution of FUV size as a function of stellar mass. Figure \ref{mass_fuvsize}(a) shows the SDSS master sample, while Figure \ref{mass_fuvsize}(b) plots only the ESF-candidate sample. A strong trend is apparent in both panels, with more massive galaxies having larger UV diameters. Notably, the \esf\ are consistent with this trend. The median FUV FWHM residual of the \esf\ from a linear fit to the ESF-candidate sample is now only $0.08\pm0.04$ dex, i.e., at fixed stellar mass, the \esf\ are $\sim20\%$ larger. This median UV size residual is \emph{smaller} than in the previous comparison in Figure \ref{fuvr_fuvsize}(b), which did not take into account the strong trend with stellar mass. Again, the existence of LINER galaxies with similarly large UV sizes strengthens our point that such UV-extended galaxies comprise a non-negligible fraction of massive GV ETGs (see below).

To summarize, we conclude that the stellar masses and UV sizes of the 19 \emph{HST} \esf\ are consistent with a much larger sample of ESF-candidate objects. This analysis lends support to our claim that the \esf\ are only a small subset of a larger population of galaxies with similar properties.

\subsubsection{The Contribution of \esf\ to the GV}\label{contamination}

Investigating a sample with weak or marginally detected properties is always challenging because of the inherent inhomogeneities present within it due to measurement error or misidentification. The ESF-candidate sample is no exception, and properly screening its members is necessary to obtain useful results. Indeed, prior to formulating their final \hst program, \sr\ visually inspected their $\sim60$ candidate objects for possible contamination due to nearby companions, tidal disturbances, or post-starburst galaxies and wound up discarding half of them on these accounts. We repeated this exercise on their $\sim60$ objects and found 20 contaminants, or $1/3$ of the sample. Most of the contamination is due to nearby UV-bright objects that in some cases induce a large ($\ga3\arcsec$) offset between the SDSS and GALEX positions. Five contaminants have post-starburst spectra and appear rather blue in optical color; these were eliminated by \sr\ because their UV emission is obviously due to SF. In addition to the 1/3 of the sample excluded by our new visual inspections, \hst UV imaging revealed that about 1/3 of the actually imaged galaxies had UV morphologies unlike the \esf; their UV emission is either compact or on small scales (Section \ref{data} and Paper I). Taken together, these cuts imply that almost half of the galaxies in the initial sample of $\sim60$ are not bona fide \esf. Assuming that this fraction is representative of the larger ESF-candidate sample, the estimated number of true \esf\ is reduced from $\sim1200$ to $\sim600$.

A key question is whether the ESF-candidate sample represents the typical galaxy evolving through the GV. However, determining this requires some care. Because the \esf\ and the ESF-candidates have high stellar masses ($\ga10^{10.5}M_\odot$), we can only make statements about the \emph{massive} end of the GV. We define the massive end of the GV in \fuvr color-mass space to be bounded by $4<$ \fuvr $<5.3$ and $10.3<\log M_*/M_\odot <11.2$. This mass range spans most of the \esf\ and contains $\sim5200$ galaxies in the SDSS master sample. To avoid potential contamination due to dust-reddened, star-forming galaxies, we further exclude highly inclined galaxies with axis ratio $b/a\le0.5$. This is a significant cut that removes $\sim1800$ galaxies. This cut is motivated by the observation that nearly all of the \esf\ tend to be nearly face-on, with $b/a>0.5$ (Table \ref{table_prop}). To be consistent with the cuts made to identify massive GV galaxies, we cut the ESF-candidate sample to have the same range in stellar mass, \fuvr color, and axis ratio. This results in $670$ galaxies, and implies that the relative fraction of ESF-candidates among massive GV galaxies in this UV color-mass bin is $670/3400\approx20\%$. Exact numbers are presented in Table \ref{gvcounts} for reference.

Note that this quoted fraction has not been corrected for contamination by superposed companions, post-starburst galaxies, etc., as discussed above. If such contamination is random, then the resulting fraction remains unchanged. However, if some forms of contamination affect the ESF-candidate sample more than the general GV population, then the fraction can be smaller. For example, post-starburst galaxies ought to be excluded from the ESF-candidates but not the general GV because they are truly transitioning from blue to red but are not \esf. We have attempted to estimate this ``differential contamination'' by applying cuts to the ESF-candidate sample that mimic the results of the visual inspection described earlier. Specifically, contaminants are excluded if they satisfy any of the following conditions: (1) H$\delta_A>1.65$ to remove post-starbursts, (2) FUV-to-optical size ratio $<2.15$ to remove galaxies without extended SF, or (3) GALEX-SDSS separation $>2\farcs2$ to remove obvious superpositions. Applying these cuts lowers the ESF-candidate fraction to $\approx13\%$, i.e., contamination may reduce the fraction by $\approx30\%$.

\begin{deluxetable}{lc} 
\tablewidth{0in}
\tablecaption{Contribution of ESF-ETGs to the GV \label{gvcounts}}

\tablehead{\colhead{Cut} & \colhead{SDSS Master Sample}}

\startdata
Initial Sample & 31763 (100\%) \\
Massive GV\tablenotemark{a} & 5199 (16.4\%) \\
Face-on ($b/a > 0.5$) & 3389 (10.7\%) \\
\hline
ESF-Candidates\tablenotemark{b} & 670 (2.1\%)

\enddata

\tablenotetext{a}{Galaxies with $10.3<\log M_*/M_\odot<11.2$ and $4<\mathrm{FUV}-r<5.3$.}
\tablenotetext{b}{ESF-candidates (Table \ref{mycounts}) located in the massive GV.}

\end{deluxetable}

By defining a suitable ESF-candidate sample, we have shown that many other objects exist with similar properties (\fuvr color, \halpha EW, stellar mass, and UV size) to the original \esf. Within these selection cuts, ESF-candidates classified as LINERs have similar properties as well. This bolsters our belief that \esf\ are accompanied by weak LINER-like emission. We have therefore shown that the 19 \esf\ studied in this paper are part of a larger population of similar galaxies. Moreover, we have shown that extended SF in ETGs is not a rare phenomenon among massive GV galaxies. 

\subsubsection{Constraining the Timescale of Recent SF in the \esf}\label{gvtime}

In the previous section we showed that a modest fraction ($13\%$) of massive GV galaxies are likely to be \esf. However, this by itself does not constrain the star formation \emph{timescale} of the \esf. This quantity is important as it will be used later to identify plausible SF histories for the \esf, in particular focusing on the recent SF traced by the UV. In this section we estimate the contribution of rapid, bursty SF as triggered by a minor merger to the fraction of ETGs in the GV. The goal of this analysis is to determine if such bursty SF is sufficient to explain the abundance of ESF-candidates in the GV identified in the previous section. 

The expected number of ETGs found in the GV due to minor-merger-driven SF is calculated as follows. We assume the progenitors are ETGs in the UV-optical red sequence (\fuvr$>6$) or not detected at all in FUV. This color cut is made on the assumption that ESF-ETG progenitors were initially quiescent and not star-forming and that the subsequent, rejuvenated SF moved them back into the GV. The progenitors are restricted to galaxies with (1) $10.3<\log M_*/M_\odot <11.2$, (2) concentration $C>2.5$, and (3) axis ratio $b/a>0.5$. These cuts are made to be consistent with the ESF-candidates, which are the presumed descendants. This gives 11875 galaxies. 

The minor merger rate is estimated using the theoretical predictions given in \citet{kaviraj09}. Specifically, a minor merger rate of 0.03 $\mathrm{Gyr}\,^{-1}$ (1:10 mass ratio) is assumed. In addition we estimate the time spent in the GV due to enhanced SF to be $\sim750$ Myr. This estimate assumes three pericentric passages of the satellite, each causing an enhancement of the SF rate (and bluer \fuvr colors) lasting for $\sim250$ Myr \citep{kaviraj09,peirani10}. Note that the final coalescence is not included in this estimate because the resulting SF is expected to be centrally concentrated and hence unlike the SF seen in the \esf. Combining these values, we estimate $\approx270$ ETGs to be present in the GV due to (ring-like) minor-merger-driven SF. Compared with the 670 ESF-candidates identified in the previous section, we see that bursty SF resulting from minor mergers can only account for $\sim40\%$ of ESF-candidates. It is interesting to note that this fraction is not too different from the $\sim30\%$ differential contamination presented in the previous section. In other words, the contaminating galaxies are quite possibly a subset of the minor-merger-driven SF galaxies. Note that the above discussion is meant to constrain the role of bursty \emph{star formation} due to minor mergers; we are not requiring that the \emph{gas} fueling the observed SF cannot have come from minor mergers.

It should be noted that the 40\% estimate is subject to uncertainties in both the merger rate and GV residence time. In particular, (minor) merger rates are uncertain at the factor of $\sim2$ level at least \citep[e.g.,][]{hopkins10,lotz11}. Also, the GV residence time is sensitive to the duration of each burst of SF and the number of passages, which depend on the merger geometry, mass ratio, etc.~\citep[e.g.,][]{cox08,peirani10}. It is difficult to quantify such uncertainties, and the 40\% contribution estimated above should be viewed as a rough estimate at best. Thus we tentatively suggest that bursty SF is unlikely to be the dominant cause of the SF in the \esf, though it cannot be firmly ruled out by this analysis. However, Section \ref{discussion} below highlights other evidence that indicates that the SF in the majority of \esf\ is likely due to more gradual processes.


\section{HOW DO \esf\ MOVE THROUGH THE GREEN VALLEY?}\label{discussion}

Given the photometric and spectroscopic results presented for the \esf, along with the selection and examination of the ESF-candidates, we are now in a stronger position to interpret the recent SF observed in the \esf. The main question motivating this study is whether the \esf\ are fading out or are rejuvenated. This question is directly tied to how these galaxies are evolving through the GV. To provide a framework for discussion, we present three possible evolutionary scenarios to explain the SF in the \esf: (1) rapid quenching, (2) disk rejuvenation (SF in a pre-existing disk), and (3) gradual decline. We evaluate the plausibility of each scenario in light of the results in this paper and Paper I. The present discussion supplements the close examination in Paper I of the origin of the gas fueling the extended SF in the \esf.

\emph{(1) Rapid quenching.} In this scenario, SF is quickly truncated as a result of a (gas-rich) major merger that triggers a rapid starburst and/or AGN. Centrally concentrated SF is expected to occur in galaxies undergoing such an event due to the funneling of gas toward the center during the merger \citep[e.g.,][]{hernquist95,hopkins06}. The gas can also power an AGN that ejects any remaining gas out of the galaxy \citep{dimatteo05}. The timescale for SF is expected to be short \citep[a few 100 Myr;][]{dimatteo05,hopkins08}, owing to the strong AGN feedback that effectively quenches any remaining SF. The resulting remnant is predicted to have spheroidal morphology. Note that this scenario applies equally well to a galaxy that is moving from the blue cloud to the red sequence for the first time or has been rejuvenated back into the GV.

Though it is possible that the \esf\ have undergone rapid quenching in the past, the \emph{recent} SF observed is unlikely to be related to the end phase of such an event. Their optical morphologies, while bulge-dominated, include a disk component, are undisturbed, and do not reveal any signs of recent mergers or interactions with nearby companions (SR17 is an exception; Paper I). In addition, we find that the stellar populations at the centers of the \esf\ are uniformly old, arguing against centrally concentrated recent SF. There does exist evidence for rapid quenching of SF via major mergers in certain kinds of galaxies, notably post-starburst galaxies. However, the spectra of such objects show signatures of a few 100-Myr old stellar population superimposed on old stars \citep[e.g.,][]{quintero04}, and many are morphologically disturbed. In contrast, none of the SDSS spectra of the \esf\ displays the strong Balmer lines characteristic of post-starbursts, as expected from how they were selected. We note that only five of the original 60 \sr\ candidates were excluded because they appeared to be post-starbursts based on their large H$\delta_A$ values; these may be the post-major merger objects. The lack of central SF in the \esf\ is not too surprising given the \sr\ selection criteria, which selected galaxies without detectable emission in the centers yet had blue \fuvr colors. This unique combination of parameters is thus a useful method of finding GV galaxies dominated by \emph{extended} UV emission.

Based on their undisturbed morphologies and lack of centrally concentrated SF, we conclude that the \esf\ have not experienced recent rapid quenching and that their SF is the result of some other, less violent process. We are not implying, however, that such a scenario is impossible for at least some GV galaxies. Rapid quenching is seriously considered as the main channel for some galaxies to evolve onto the red sequence \citep{faber07,martin07}.

\emph{(2) Disk rejuvenation.} In this scenario recent SF takes place in an \emph{existing} (formerly quiescent) stellar disk, such as those found in S0 galaxies. The newly formed stars cause the \fuvr color to become bluer. Because the new young stars form on top of the existing old stellar population, galaxies undergoing disk rejuvenation would be expected to have similar colors to the outer colors seen in the \esf\ in our sample. In the nearby universe, this phenomenon is exemplified by the discovery of extended-UV (XUV) disks \citep[e.g.,][]{thilker05,thilker07}. XUV-disk galaxies are characterized by extended UV emission located beyond the SF threshold. In most cases, this translates to SF being located beyond the optical extent of the galaxy, though in some objects the XUV emission is found \emph{within} the optical extent. XUV-disks have been found in galaxies of all Hubble types, notably even in E/S0 galaxies \citep{lemonias11,moffett12}. Thus it is possible that the \esf\ are examples of early-type XUV-disks, especially since their SF occurs primarily in the outskirts of the optical disks (Figure \ref{plot_prof}; Paper I). 

\citet{thilker07} adopted a threshold UV surface brightness of $\mu_\mathrm{FUV}=27.25$ AB mag arcsec$^{-2}$ to characterize XUV disks, which corresponds to a threshold SF surface density of $\Sigma_\mathrm{SFR}=3\times10^{-4}\, M_\odot\,\mathrm{yr\,^{-1}}\,\mathrm{kpc^{-2}}$. According to that criterion, our ESF-ETG UV disks hover near that level, as may be seen in Figure \ref{plot_prof}, thus qualifying them as XUV disks or very near that level. Indeed, a study of the space density of XUV-disk galaxies by \citet{lemonias11} has shown that XUV-disks are more common among red sequence and GV galaxies. In one respect, however, our galaxies differ strongly in that their outer \fuvr colors are a full 1.5-2.5 mag bluer than their integrated \fuvr colors. In \citeauthor{lemonias11}, by contrast, outer \fuvr colors correlate closely with integrated \fuvr colors, and there are no blue outer disks at all among their red galaxies (their Figure 12). Thus, it is not clear that they are detecting the same phenomenon that we are.

Different mechanisms can deliver gas into the outer regions of a galaxy and trigger disk rejuvenation. \citet{lemonias11} proposed that cold gas accretion from the intergalactic medium (IGM) is responsible for triggering SF in the XUV-disks. This picture is a possible manifestation of late-time cold mode accretion predicted in simulations \citep{keres09}. In addition, minor, gas-rich mergers can also trigger recent SF in ETGs \citep{kaviraj09}, perhaps leading to disk rejuvenation in addition to nuclear SF. 

Disk rejuvenation may be a viable explanation for the SF seen in at least some of the \esf. All the \esf\ in the sample are classified as S0 or later (Paper I), so the presence of extended UV emission is not unreasonable given the literature results discussed above. As discussed in Section \ref{sps_ext}, the outer disks of the \esf\ contain a significant underlying old stellar population upon which new stars have formed. However, any disk rejuvenation cannot have been triggered recently or very rapidly. For one thing, the highly symmetric UV morphologies of the \esf\ would require any gas to be resident in the galaxies for a few rotation periods, which in the outer parts is likely to exceed $\sim1$ Gyr. IGM accretion is consistent with this picture as long as it occurs smoothly on extended timescales, allowing the gas to settle into an axisymmetric disk before forming stars. 

In addition, as Figure \ref{models} indicates, a burst superimposed on an old stellar population reddens too quickly. We roughly estimated in Section \ref{gvtime} that bursty SF due to minor mergers produces less than half the number of ESF-candidates identified in the GV (Section \ref{contamination}), tentatively suggesting that minor mergers are not the dominant cause of the SF in the \esf. Moreover, the SF histories predicted from minor merger simulations are bursty with typical timescales of a few to several 100 Myr \citep{cox08,kaviraj09}. Also, the newly formed stars tend to be centrally concentrated \citep{cox08,peirani10}, in contrast with the red colors seen in the bulges of the \esf\ (Figure \ref{plot_prof}). Taken together, these points suggest that disk rejuvenation is only viable if (1) the gas has not been recently accreted and (2) the resulting SF has been ongoing for an extended period of time. 

In a study of the recent SF histories of SDSS galaxies, \citet{kauffmann06} hypothesized that galaxies above a transitional surface mass density experience intense bursty SF that consumes infalling gas on timescales of $\sim1$ Gyr. Such (rejuvenated) SF is likely to drive galaxies from the red sequence into the green valley. The morphological and photometric data of the \esf\ can provide useful constraints on this model. In particular, assuming a rotation speed of 200 km~s$^{-1}$ at a radius of 10 kpc, and hence a rotation period of 300 Myr, recently infalling gas would have experienced $\sim3$ rotation periods in 1 Gyr. This is likely sufficient time for the gas to settle into an axisymmetric disk. The average galaxy would be observed halfway through this period, and one would expect to see a distribution of morphologies corresponding to various stages of gas rearrangement. However, we do not see such diversity in the UV morphologies of the \esf. In addition, stellar population analysis of the outer disks (Section \ref{sps_ext}) indicates that SF in the outer parts of the \esf\ takes place on timescales longer than 1 Gyr. Also, the similarity between the outer colors of the  \esf\ and massive blue cloud galaxies [where SF proceeds on Gyr timescales \citep{noeske07}] points to a longer gas consumption timescale than that suggested by \citet{kauffmann06}.

\emph{(3) Gradual decline.} In this scenario, a galaxy's \emph{original} SF gradually shuts down as its existing gas supply is used to form stars \citep{noeske07}. No mergers or interactions are required to explain the recent SF, and the galaxy's original (disky) morphology is preserved. This model reflects the general expectation that most galaxies are evolving from blue to red for the first time. Gradual decline implies that SF is steadily decreasing over time, and hence that there exists an underlying population of older stars. 

While the timescale for evolution from blue to red is generally believed to be fairly rapid \citep[$\sim1$ Gyr;][]{martin07}, it is certainly not expected that \emph{all} galaxies need transition through the GV in such a rapid manner. Indeed, local observations have shown the existence of a population of galaxies in the GV, primarily S0s, whose cold gas properties imply a much longer GV residence time \citep[e.g.,][]{vandriel91,noordermeer05,cortese09}. Specifically, radio observations of S0s have revealed that some harbor large \ion{H}{1} reservoirs containing $\sim10^{9}\,M_\odot$ of gas or more. The gas surface density in these reservoirs, $\Sigma_\mathrm{gas}\sim \mathrm{few}\, M_\odot\,\mathrm{pc}^{-2}$, fall within the \ion{H}{1}-to-H$_2$ transition regime, where the SF rate declines non-linearly with $\Sigma_\mathrm{gas}$ \citep{krumholz09}. Specifically, the SF rate is predicted to be $\sim10$ times lower than in normal spiral galaxies, i.e., $\sim0.1\,M_\odot\,\mathrm{yr}\,^{-1}$. These results imply a gas consumption time of up to 10 Gyr in these gas-rich S0s. In other words, there likely exists a subpopulation of GV galaxies whose SF can persist at a very low level for a very long time.   

We find this gradual decline scenario to be another possible explanation for the recent SF seen in the \esf. Their undisturbed and symmetric morphologies indicate a lack of recent mergers or interactions. The blue outer disks plus red bulges are a possible extension of morphologies of galaxies slightly later along the Hubble sequence, which sometimes show a hole in their central gas distributions and a lack of young stars in their bulges \citep{pahre04,gildepaz07}. Moreover, the \ion{H}{1} distributions of the S0 galaxies described above are typically ring-like with central depressions in gas surface density \citep[e.g.,][]{vandriel91,delrio04,noordermeer05}. Moreover, GALEX imaging of some of these S0s show UV rings reminiscent of those seen in the \esf\ \citep{cortese09}. Thus it is conceivable that the \esf\ harbor similar reservoirs of low-surface-density cold gas. Obtaining radio observations of the \esf\ would be extremely useful in characterizing the amount and distribution of cold gas in these objects.

Finally, we showed in Section \ref{sps_ext} that the stellar populations in the UV rings contain a significant contribution from old stars, and that their SF histories are consistent with fading $\tau$-models. Assuming that the fading model applies to \esf\ in general, and because fading is considered an important mode of evolution through the GV, one would expect to find a significant fraction of GV galaxies with ESF-like properties. Section \ref{contamination} showed that $\approx13\%$ of massive galaxies in the bluest part of the GV are ESF-candidates. Though not a majority in this region, ESF-candidates nonetheless make up a non-negligible portion of it, and we cannot rule out fading for at least a subset of these objects. We note that gas-rich GV galaxies comprise $\sim12\%$ of the total GV sample presented in \citet{cortese09}. The close agreement between their fraction and the $13\%$ ESF-ETG GV fraction found here is intriguing and indicates that gradual decline of SF on Gyr timescales should not be discounted as an important evolutionary path through the GV for some objects.

To summarize, only certain models for the SF in the \esf\ survive comparison with the results presented in this paper. The viable models are the slow decline of the original SF and/or (non-bursty) disk rejuvenation. \sr\ and Paper I reach similar conclusions based on the integrated properties of the \esf. In particular, both works found that the decline of the original SF is possible in a subset of the \esf, and we come to a similar conclusion based on stellar population analysis of the outer disks of the \esf.

\sr\ also argue that the SF in some of the \esf\ is most consistent with rejuvenation because of their lower intrinsic dust content and larger-than-average UV sizes relative to the entire GV. While \sr\ supported rejuvenated SF, they were unable to constrain whether the necessary gas was accreted via mergers or directly from the IGM. We find that mergers (in the rapid quenching model) are unlikely to produce the observed morphologies and lack of central SF in the \esf, but IGM accretion is permissible as long as it does not disrupt the galaxy and the resulting SF is not too recent and bursty. Paper I reaches a similar conclusion, finding that $\sim55\%$ of the \esf\ have UV and optical morphologies consistent with smooth IGM accretion. Rejuvenation is permitted as long as the new stars began forming long ago in a gradual manner. We have not been able to think of tests that could tell the difference, definitively, between this kind of ``extended rejuvenation'' and a simple decline of the existing SF.

\section{SUMMARY AND CONCLUSIONS}\label{conclusions}

This paper complements the work of \sr\ and Paper I. Using high-resolution UV imaging from \emph{HST} and optical photometry from SDSS, we measure surface brightness and UV-optical color profiles of 19 \esf\ selected from a cross-matched SDSS/GALEX catalog. These galaxies were selected to have high concentration, blue \fuvr color, and no detectable emission lines within the SDSS fiber. This combination of selection criteria yielded a sample dominated by objects with \emph{extended} UV emission. Our main results are summarized below.

\begin{enumerate}
	
	\item The UV-optical color profiles of the \esf\ all show red centers and blue outer parts, as expected given how they were selected. The central \fuvr colors are consistent with red sequence galaxies (\fuvr$\ga 6$), implying that the observed central UV emission is unlikely due to recently formed stars. By contrast, the outer \fuvr colors are all quite blue (\fuvr$\sim2-4$), which is similar to star-forming, blue cloud galaxies. Blue colors together with well-formed UV structures in the regions of extended UV emission point to recent SF in the outer regions of these galaxies.
	
	\item A comparison of the outer colors of the \esf\ with blue cloud galaxies in different mass ranges shows that the stellar populations in the outer disks of the \esf\ contain many old stars. Additionally, their outer colors are consistent with gradually declining $\tau$-models that indicate old stellar population ages, as well as with a ``double-$\tau$'' model representing a prolonged rejuvenation event. The colors are also marginally consistent with a recent-burst model. However, we roughly estimate that a bursty SF history triggered by minor mergers underpredicts the number of ESF-candidates we find (Point 4 below) by a factor of $\sim2$. 
	
	\item Based on stacked SDSS spectra, we determine a mean light-weighted age of $6.8^{+2.1}_{-1.7}$ Gyr for the centers of the \esf, consistent with their red \fuvr colors. We find evidence for weak LINER-like emission (i.e., [\ion{O}{2}] and [\ion{N}{2}] emission) in the central regions of the \esf. If the emission is due to weak AGN activity, this suggests that AGN feedback may have played a role in quenching SF, particularly in the centers of these objects. If instead the LINER emission is caused by gas ionization due to old stars, then this strengthens our conclusion that the centers are dominated by evolved stellar populations.
	
	\item By using a larger, updated SDSS/GALEX catalog, we find $\sim600$ ESF-candidate galaxies that have similar spectral and structural properties to the \esf. The key change in sample selection that yielded a much larger number of ESF-candidates was admitting all H$\alpha$-weak objects, including LINERs. We find that the ESF-candidates and true \esf\ have similar distributions in \fuvr color, H$\alpha$ EW, UV size, and stellar mass. A rough estimate suggests that \esf\ comprise $\approx13\%$ of massive GV galaxies in their UV color-mass range.
	
	\item We present three possible scenarios to explain the recent SF in the \esf. Based on the results presented, we are able to rule out rapid quenching following a merger-induced starburst. However, our results are consistent with disk rejuvenation (via smooth IGM accretion) and/or the gradual decline of the original SF. Both scenarios can explain the regular morphologies of the \esf\ and the presence of underlying stars in the regions of recent SF. The similarity of some \esf\ to local, gas-rich S0s in the GV suggests that they are among a class of GV galaxies that may take Gyrs to fully quench and reach the red sequence. 
		
\end{enumerate}

The discovery of recent SF in ETGs has challenged traditional explanations of the evolution of these galaxies. Our understanding has benefitted greatly from all-sky surveys in the optical and, crucially, in the UV. The ability of the UV to detect even faint signatures of SF in seemingly quiescent objects highlights the richness and diversity of ETGs and raises many important questions, some of which we examined in this work. Our study of the \esf\ represents an initial attempt to constrain the evolution of galaxies in the GV. Future work will systematically examine the properties and numbers of a wider range of galaxies to study the mass-dependent evolution of galaxies through the GV. 

\acknowledgements

We warmly thank the anonymous referee for a careful and thorough report that has improved the clarity of this work. We thank David Koo and members of the DEEP collaboration at UCSC for stimulating discussions. J.~J.~F. acknowledges financial support from a UCSC Regent's Fellowship and \emph{HST} grant GO-11175. R.~M.~R. acknowledges support from \emph{HST} grant GO-11158. 

Based on observations made with the NASA/ESA Hubble Space Telescope, obtained at the Space Telescope Science Institute, which is operated by the Association of Universities for Research in Astronomy, Inc., under NASA contract NAS 5-26555. 

Funding for the SDSS and SDSS-II has been provided by the Alfred P. Sloan Foundation, the Participating Institutions, the National Science Foundation, the U.S. Department of Energy, the National Aeronautics and Space Administration, the Japanese Monbukagakusho, the Max Planck Society, and the Higher Education Funding Council for England. The SDSS Web Site is http://www.sdss.org/.

The SDSS is managed by the Astrophysical Research Consortium for the Participating Institutions. The Participating Institutions are the American Museum of Natural History, Astrophysical Institute Potsdam, University of Basel, University of Cambridge, Case Western Reserve University, University of Chicago, Drexel University, Fermilab, the Institute for Advanced Study, the Japan Participation Group, Johns Hopkins University, the Joint Institute for Nuclear Astrophysics, the Kavli Institute for Particle Astrophysics and Cosmology, the Korean Scientist Group, the Chinese Academy of Sciences (LAMOST), Los Alamos National Laboratory, the Max-Planck-Institute for Astronomy (MPIA), the Max-Planck-Institute for Astrophysics (MPA), New Mexico State University, Ohio State University, University of Pittsburgh, University of Portsmouth, Princeton University, the United States Naval Observatory, and the University of Washington.

GALEX (Galaxy Evolution Explorer) is a NASA Small Explorer, launched in 2003 April. We gratefully acknowledge NASA's support for construction, operation, and science analysis for the GALEX mission, developed in cooperation with the Centre National d'\'Etudes Spatiales of France and the Korean Ministry of Science and Technology.

{\it Facilities:} \facility{GALEX}, \facility{\emph{HST} (ACS)}, \facility{Sloan}



\end{document}